\documentclass[11pt,fleqn]{article}
\usepackage{amssymb,latexsym,amsmath,amsfonts,graphicx}

    \advance\textheight by \topskip

    \numberwithin{equation}{section}

    \def\tr{{\rm tr \,}}
    \def\Re{{\rm Re \,}}
    \def\Im{{\rm Im \,}}
    \def\Ai{{\rm Ai \,}}
    \def\I{{\rm I \,}}
    \def\II{{\rm II \,}}
    \def\III{{\rm III \,}}
    \def\IV{{\rm IV \,}}
    \def\bigO{{\cal O}}

    \newtheorem{theorem}{Theorem}[section]
    \newtheorem{lemma}[theorem]{Lemma}
    \newtheorem{corollary}[theorem]{Corollary}
    \newtheorem{proposition}[theorem]{Proposition}
    
    \newtheorem{Definition}[theorem]{Definition}
    
    \newtheorem{Remark}[theorem]{Remark}
    \newenvironment{remark}{\begin{Remark}\rm}{\end{Remark}}
    \newtheorem{Example}[theorem]{Example}
    \newenvironment{example}{\begin{Example}\rm}{\end{Example}}
    \newtheorem{Assumptions}[theorem]{Assumptions}
    \newenvironment{assumptions}{\begin{Assumptions}\rm}{\end{Assumptions}}

    \newenvironment{proof}%
    {\rm \trivlist \item[\hskip \labelsep{\bf Proof. }]}%
    {\hspace*{\fill}$\Box$\endtrivlist}
    \newenvironment{varproof}%
    {\rm \trivlist \item[\hskip \labelsep{\bf Proof}]}%
    {\hspace*{\fill}$\Box$\endtrivlist}

    \newcommand{\sgn}{{\operatorname{sgn}}}
    \newcommand{\Arg}{{\operatorname{Arg}}}
    
    \newcommand\PVint{\mathop{\setbox0\hbox{$\displaystyle\intop$}%
        \hskip0.2\wd0%
        \vcenter{\hrule width0.6\wd0height0.5pt depth0.5pt}%
        \hskip-0.8\wd0%
        }\mskip-\thinmuskip\intop\nolimits}

\begin{document}
\title{Universality of a double scaling limit near singular edge points in
random matrix models}
\author{T. Claeys and M. Vanlessen}

\maketitle

\begin{abstract}
    We consider unitary random matrix ensembles $Z_{n,s,t}^{-1}e^{-n\,\tr V_{s,t}(M)}dM$ on the space
    of Hermitian $n\times n$ matrices $M$,
    where the confining potential $V_{s,t}$ is such that
    the limiting mean density of eigenvalues (as $n\to\infty$ and $s,t\to
    0$) vanishes like a power $5/2$ at a (singular) endpoint of its support. The main
    purpose of this paper is to prove universality of the
    eigenvalue correlation kernel in a double scaling
    limit. The limiting kernel is built out of functions
    associated with a special solution of the $P_I^2$ equation,
    which is a fourth order analogue of the Painlev\'e I equation.
    In order to prove our result, we use the well-known connection between the
    eigenvalue correlation kernel and the Riemann-Hilbert (RH) problem for
    orthogonal polynomials, together with the Deift/Zhou
    steepest descent method to analyze the RH problem
    asymptotically. The key step in the asymptotic analysis will be the
    construction of a parametrix near the singular endpoint, for which we use
    the model RH problem for the special solution of the $P_I^2$
    equation.

    In addition, the RH method allows us to determine
    the asymptotics (in a double scaling limit) of the
    recurrence coefficients of the orthogonal polynomials with respect to the varying weights
    $e^{-nV_{s,t}}$ on $\mathbb R$. The
    special solution of the $P_I^2$ equation pops up in the $n^{-2/7}$-term
    of the asymptotics.
\end{abstract}

\section{Introduction and statement of results}

\subsection{Unitary random matrix ensembles}
    \label{subsection: unitary ensembles}

On the space $\mathcal H_n$ of Hermitian $n\times n$ matrices $M$, we consider
for $n\in\mathbb N$ and $s,t\in\mathbb{R}$ the unitary random matrix ensemble,
\begin{equation}\label{random matrix model}
\frac{1}{Z_{n,s,t}}e^{-n\,\tr V_{s,t}(M)}dM.
\end{equation}
Here, $Z_{n,s,t}$ is a normalization constant and the confining
potential $V_{s,t}$ is a real analytic function, depending on two
parameters $s,t\in\mathbb R$, satisfying the asymptotic condition,
\begin{equation} \label{conditionV}
    \lim_{x\to\pm\infty} \frac{V_{s,t}(x)}{\log(x^2+1)} = +\infty,
        \qquad \mbox{uniformly for $s,t\in[-\delta_0, \delta_0]$ for some $\delta_0>0$.}
\end{equation}
Then,
\[
    Z_{n,s,t}=\int_{\mathcal H_n}e^{-n\,\tr V_{s,t}(M)}dM
\]
is convergent as $n\to\infty$ so that the random matrix model is well-defined.

It is well-known, see e.g.\ \cite{Mehta}, that an important role in the study
of the unitary random matrix ensemble (\ref{random matrix model}) is played by
the following scalar 2-point (correlation) kernel,
\begin{equation} \label{kernel}
    K_n^{(s,t)}(x,y)=
        e^{-\frac{n}{2}V_{s,t}(x)} e^{-\frac{n}{2}V_{s,t}(y)}
        \sum_{k=0}^{n-1} p_k^{(n,s,t)}(x) p_k^{(n,s,t)}(y),
\end{equation}
constructed out of the orthonormal polynomials
\[
    p_k^{(n,s,t)}(x)=\kappa_k^{(n,s,t)} x^k + \cdots,
    \qquad\qquad \mbox{$\kappa_k^{(n,s,t)}>0$,}
\]
with respect to the varying weights $e^{-nV_{s,t}}$ on $\mathbb R$.
Indeed, the correlations between the eigenvalues of $M$ can be
written in terms of the correlation kernel. More precisely, the
$m$-point correlation function $\mathcal R_{n,m}^{(s,t)}$ satisfies
\cite{Mehta},
\begin{equation}
    \mathcal R_{n,m}^{(s,t)}(x_1,\ldots
    ,x_m)=\det\left(K_n^{(s,t)}(x_i,x_j)\right)_{1\leq i,j\leq m}.
\end{equation}
Further, the limiting mean eigenvalue distribution $\mu_{s,t}$ has a density
$\rho_{s,t}$ which can be retrieved from the correlation kernel as follows,
\begin{equation}\label{psit-kernel}
    \rho_{s,t}(x)=\lim_{n\to\infty}\frac{1}{n}K_n^{(s,t)}(x,x).
\end{equation}

\medskip

The limiting mean eigenvalue distribution $\mu_{s,t}$ equals \cite{DKMVZ2} the
equilibrium measure in external field $V_{s,t}$. This is the unique measure
minimizing the logarithmic energy \cite{SaTo}
\begin{equation}\label{definition: energy}
    I_{V_{s,t}}(\mu)=
        \iint \log \frac{1}{|x-y|}d\mu(x)d\mu(y)
        +\int V_{s,t}(y)d\mu(y),
\end{equation}
among all probability measures $\mu$ on $\mathbb R$. Furthermore, there exists
a real analytic function $q_{s,t}$, such that \cite{DKM},
\begin{equation}\label{definition: qst}
    \rho_{s,t}(x)=\frac{1}{\pi}\sqrt{q_{s,t}^-(x)},
\end{equation}
where $q_{s,t}^-$ denotes the negative part of $q_{s,t}$, i.e.
$q_{s,t}=q_{s,t}^+-q_{s,t}^-$, with $q_{s,t}^\pm\geq 0$ and
$q_{s,t}^+q_{s,t}^-=0$. Due to condition (\ref{conditionV}) we have
that $q_{s,t}(x)\to +\infty$ as $x\to\pm\infty$, so that $\mu_{s,t}$
is supported on a finite union of intervals, which we denote by
$\mathbb S_{s,t}$. It is known \cite{SaTo} that the equilibrium
measure $\mu_{s,t}$ satisfies the following Euler-Lagrange
variational conditions: there exists a constant
$\kappa_{s,t}\in\mathbb{R}$ such that
\begin{align}
    \label{variationalcondition:must-equality1}
    & 2\int \log |x-u|d\mu_{s,t}(u)-V_{s,t}(x)=\kappa_{s,t},
        &\mbox{for $x\in \mathbb S_{s,t}$,}
    \\[1ex]
    \label{variationalcondition:must-inequality1}
    & 2\int \log |x-u|d\mu_{s,t}(u)-V_{s,t}(x)\leq \kappa_{s,t},
        &\mbox{for $x\in \mathbb R\setminus \mathbb S_{s,t}$.}
\end{align}

The external field $V_{s,t}$ is called regular if strict inequality in
(\ref{variationalcondition:must-inequality1}) holds, if the density
$\rho_{s,t}$ does not vanish in the interior of the support $\mathbb S_{s,t}$,
and if $q_{s,t}$ has a simple zero at each of the endpoints of the support
$\mathbb S_{s,t}$. If one of these conditions is not valid, $V_{s,t}$ is called
singular. The singular points $x^*$ are classified as follows, see
\cite{DKMVZ2, KM}:

\begin{itemize}
    \item[(i)] $x^*\in\mathbb{R}\setminus \mathbb S_{s,t}$ is a type $\I$
        singular point if equality in (\ref{variationalcondition:must-inequality1})
        holds. Then, $x^*$ is a zero of $q_{s,t}^+$ of multiplicity $4m$ with $m\in\mathbb N$.
    \item[(ii)] $x^*\in\mathbb S_{s,t}$ is a type II singular point if it is an
        interior point of $\mathbb S_{s,t}$ where the equilibrium density
        $\rho_{s,t}$ vanishes. Then, $x^*$ is a zero of $q_{s,t}^-$ of multiplicity $4m$.
    \item[(iii)] $x^*$ is a type III singular point if it is an endpoint
        of the support $\mathbb S_{s,t}$ and a zero of $q_{s,t}$ of multiplicity larger than
        one. Then, $x^*$ is a zero of $q_{s,t}$ of multiplicity $4m+1$,
        which means that $\rho_{s,t}(x)\sim c|x-x^*|^{(4m+1)/2}$.
\end{itemize}

In this paper, we consider external fields $V_{s,t}$ which are such that in the
critical case $s=t=0$, $V_0=V_{0,0}$ has a type III singular (edge) point $x^*$
with $m=1$, i.e.
\begin{equation}
    \rho_{0,0}(x)\sim c|x-x^*|^{5/2},\qquad\mbox{as $x\to x^*$.}
\end{equation}
Further, we take $V_{s,t}$ of the special form,
\begin{equation}
    V_{s,t}=V_0+sV_1+tV_2,
\end{equation}
where $V_1$ is an arbitrary real analytic function, while $V_2$ is real
analytic and in addition satisfies some critical condition which we will
specify in Section \ref{subsection: statement of results} below.

\subsection{Universality in random matrix theory}

Consider for now unitary random matrix ensembles $Z_n^{-1}e^{-n\,\tr
V(M)}dM$ on the space of Hermitian $n\times n$ matrices $M$. Scaling
limits of the associated correlation kernel $K_n$ show universal
behavior.

Near regular points, universality results have been established in \cite{BI1,
Deift, DKMVZ2, DKMVZ1, PS}. For example, if $x^*$ lies in the bulk of the
spectrum (i.e.\ $x^*$ is such that it lies in the interior of the support
$\mathbb S$ of the equilibrium measure in external field $V$, and such that the
equilibrium density $\rho$ does not vanish at $x^*$) there is a constant $c$
such that
\begin{equation}
    \lim_{n\to\infty} \frac{1}{cn}K_n\left(x^*+\frac{u}{cn},x^*+\frac{v}{cn}\right)
        = \frac{\sin \pi(u-v)}{\pi(u-v)}.
\end{equation}
On the other hand, if $x^*$ is a regular edge point of the spectrum
(i.e.\ $x^*$ is an endpoint of $\mathbb S$ and $\rho$ vanishes like
a square root at $x^*$), there is a constant $c$ such that
\begin{equation}
    \lim_{n\to\infty} \frac{1}{cn^{2/3}}
    K_n\left(x^*+\frac{u}{cn^{2/3}},x^*+\frac{v}{cn^{2/3}}\right)=
    \frac{\Ai(u)\Ai'(v)-\Ai(v)\Ai'(u)}{u-v},
\end{equation}
where $\Ai$ is the Airy function.

Near singular points, similar results hold. In those singular cases
it is interesting to consider double scaling limits where the
external field $V$ depends on additional parameters. In \cite{BI2,
CK, CKV, Shcherbina}, an external field $V$ was considered such that
there is a type $\II$ singular (interior) point $x^*$ with $m=1$,
i.e.
\[
    \rho(x)\sim c(x-x^*)^2,\qquad \mbox{as $x\to x^*$}.
\]
If an additional parameter is included in the external field, $V_t=V/t$, one
observes for $t$ close to $1$ the transition where two intervals in the support
of the limiting mean density of eigenvalues merge to one interval through the
critical case of a type II singular point. In the double scaling limit where
$n\to\infty$ and $t\to 1$ in such a way that $c_0n^{2/3}(t-1)\to s\in\mathbb R$
for some appropriately chosen constant $c_0$, there exists a constant $c$ such
that (for the associated correlation kernel $K_{n,t}$),
\[
    \lim \frac{1}{cn^{1/3}}K_{n,t}\left(x^*+\frac{u}{cn^{1/3}},
    x^*+\frac{v}{cn^{1/3}}\right)=K^{{\rm crit,II}}(u,v;s).
\]
Here, $K^{{\rm crit,II}}(u,v;s)$ is built out of functions
associated with the Hastings-McLeod solution \cite{HastingsMcLeod}
of the second Painlev\'e equation.

\medskip

The main purpose of this paper is to obtain, for the random matrix
models in Section \ref{subsection: unitary ensembles} above, a
similar result near the type III singular (edge) point of $V_0$ with
$m=1$. We take a double scaling limit ($n\to\infty$ and $s,t\to 0$),
and the limiting kernel $K^{{\rm crit,III}}$ will be built out of
functions which are associated with a special solution of the fourth
order analogue of the Painlev\'e I equation. The case of a type III
singular (edge) point was also studied in the Physics literature
\cite{BB,BMP}.

In addition, the techniques that we use to prove this allow us to
determine the asymptotics (in a double scaling limit) of the
recurrence coefficients in the three-term recurrence relation
satisfied by the orthogonal polynomials $p_k^{(n,s,t)}$ with respect
to the varying weights $e^{-nV_{s,t}}$ on $\mathbb R$.

\subsection{$\Phi$-functions associated with a special solution of the $P_I^2$ equation}
    \label{subsection: PI2 equation}

We consider the following differential equation for $y=y(s,t)$,
which we denote as the $P_I^2$ equation,
\begin{equation}\label{PI2}
    s=ty-\left(\frac{1}{6}y^3+\frac{1}{24}(y_s^2+2yy_{ss})
        +\frac{1}{240}y_{ssss}\right).
\end{equation}
For $t=0$, this equation is the second member in the Painlev\'e I
hierarchy \cite{KKNT,KS}. The $P_I^2$ equation has been studied for
example in \cite{BMP,Kapaev,Moore} (for $t=0$) and
\cite{CV,Dubrovin} (for general $t$). The Lax pair for the $P_I^2$
equation is the linear system of differential equations
\begin{equation}\label{differential equations}
    \frac{\partial \Psi}{\partial\zeta}=U \Psi,\qquad
    \frac{\partial \Psi}{\partial s}=W \Psi,
\end{equation}
where
\begin{align}\label{introduction: U}
    & U = \frac{1}{240}
    \begin{pmatrix}
        -4y_s \zeta-(12yy_s+y_{sss}) &
        8\zeta^2+8y\zeta+(12y^2+2y_{ss}-120t) \\[1ex]
        U_{21}
        & 4y_s \zeta+(12yy_s+y_{sss})
    \end{pmatrix},\\[3ex]
    & U_{21} =
    8\zeta^3-8y\zeta^2-(4y^2+2y_{ss}+120t)\zeta+
    (16y^3-2y_s^2+4yy_{ss}+240s),
\end{align}
and
\begin{equation}\label{introduction: W}
    W =\begin{pmatrix}
        0 & 1 \\
        \zeta-2y & 0
    \end{pmatrix}.
\end{equation}
The system of differential equations (\ref{differential
equations})--(\ref{introduction: W}) can only be solvable if
$y=y(s,t)$ is a solution to the $P_I^2$ equation (\ref{PI2}). For
different solutions $y$, we have different Lax pairs.

\medskip

We are interested in the special solution $y$ which was studied in
\cite{BMP, CV, Dubrovin}. This solution $y=y(s,t)$ is characterized
by the vanishing of its Stokes multipliers $s_1$, $s_2$, $s_5$, and
$s_6$, see \cite{Kapaev} for details. It was shown in \cite{CV} that
$y$ has no poles for real $s$ and $t$, and that it has, for fixed
$t\in\mathbb R$, the following asymptotic behavior,
\begin{equation}\label{asymptotics y}
        y(s,t)=\mp (6|s|)^{1/3}\mp \frac{1}{3}6^{2/3}t|s|^{-1/3}
            +\bigO(|s|^{-1}),
            \qquad\mbox{as $s\to\pm\infty$.}
\end{equation}
It has been shown in \cite[Appendix A]{Moore} that for $t=0$, $y$ is
uniquely determined by realness and asymptotic condition
(\ref{asymptotics y}). For general $t$ we are not aware of a similar
result although it is supported by a conjecture of Dubrovin
\cite{Dubrovin} that this should hold for general $t$. For
$s,t\in\mathbb R$, the Lax pair (\ref{differential
equations})--(\ref{introduction: W}) associated with this special
choice of $y$ has a unique
solution $\bigl(\begin{smallmatrix}\Phi_1\\
\Phi_2\end{smallmatrix}\bigr)$ for which the following limit holds,
see \cite{CV, Kapaev},
\begin{multline}\label{asymptotics Phi}
    \zeta^{\frac{1}{4}\sigma_3}
    \begin{pmatrix}
        \Phi_1(\zeta;s,t)\\
        \Phi_2(\zeta;s,t)
    \end{pmatrix}
    e^{\theta(\zeta;s,t)}\,\longrightarrow\,\frac{1}{\sqrt 2}
    \begin{pmatrix}
        1\\
        -1
    \end{pmatrix}e^{-\frac{1}{4}\pi i},
    \\
    \qquad \mbox{as $\zeta\to\infty$ with $0<\Arg\,\zeta <6\pi /7$},
\end{multline}
where $\sigma_3=\left(\begin{smallmatrix}1 & 0 \\ 0& -1
\end{smallmatrix}\right)$ denotes the third Pauli-matrix, and where $\theta$ is given by
\begin{equation}\label{definition: theta}
\theta(\zeta;s,t)=\frac{1}{105}\zeta^{7/2}-\frac{1}{3}t\zeta^{3/2}+s\zeta^{1/2}.
\end{equation}
The functions $\Phi_1$ and $\Phi_2$ will appear below in the
universal limiting correlation kernel near type $\III$ singular
(edge) points of $V_0$ with $m=1$.

\subsection{Statement of results}
    \label{subsection: statement of results}

We work under the following assumptions.
\begin{assumptions}\label{assumptions}
    \
    \begin{itemize}
    \item[(i)] We consider external fields $V_{s,t}$ of the form
        \begin{equation}\label{Vst}
            V_{s,t}=V_0+sV_1+tV_2,
        \end{equation}
        where $V_0$, $V_1$, and $V_2$ are real analytic and are such that
        there exists a $\delta_0>0$ such that the following holds
        \begin{equation} \label{conditionVbis}
            \lim_{|x|\to\infty} \frac{V_{s,t}(x)}{\log(x^2+1)} =
            +\infty, \qquad \mbox{uniformly for $s,t\in[-\delta_0, \delta_0]$.}
        \end{equation}
    \item[(ii)] $V_0$ is such that the equilibrium measure $\nu_0$
        in external field $V_0$ is supported on one single interval
        $[a,b]\subset\mathbb R$, and $b$ is a type III singular (edge) point of $V_0$ with $m=1$.
        Then, $\nu_0$ is of the form \cite{DKM},
        \begin{equation}\label{psi0h0}
            d\nu_0(x) = \frac{1}{2\pi} h_0(x)\sqrt{(b-x)(x-a)}\, \chi_{[a,b]}(x)dx,
        \end{equation}
        with $\chi_{[a,b]}$ the indicator function of the set
        $[a,b]$, and with $h_0$ real analytic and satisfying,
        \begin{equation}\label{assumptions: eq1}
            h_0(b)=h_0'(b)=0, \qquad\mbox{and}\qquad h_0''(b)>0.
        \end{equation}
        Furthermore, we assume that $V_0$ has no other singular points
        besides $b$. In particular, $a$ is a regular (edge) point and we
        then have that
        \begin{equation}\label{assumptions: eq2}
            h_0(a)>0.
        \end{equation}
    \item[(iii)] $V_2$ is such that it satisfies the critical condition
        \begin{equation}\label{conditionV2}
            \int_a^b\sqrt\frac{u-a}{b-u}V_2'(u)du=0.
        \end{equation}
    \end{itemize}
\end{assumptions}

Throughout the rest of this paper we let $\mathcal V$ be the
neighborhood of the real line where $V_0,V_1,V_2$, and $h_0$ are
analytic.

\begin{example}
    The assumptions above are valid for the particular example
    where $V_0, V_1$, and $V_2$ are given by,
    \begin{equation}
        V_0(x)=\frac{1}{20}x^4-\frac{4}{15}x^3+\frac{1}{5}x^2+\frac{8}{5}x,
            \qquad V_1(x)=x,
            \qquad V_2(x)=x^3-6x.
    \end{equation}
    Then, the equilibrium measure $\nu_0$ is supported on the
    interval $[-2,2]$ and given by
    \begin{equation}
        d\nu_0(x) =
        \frac{1}{10\pi}(x+2)^{1/2}(x-2)^{5/2}\chi_{[-2,2]}(x)dx.
    \end{equation}
    It should be noted that a type III singular (edge) point cannot
    occur when $V_0$ is a polynomial of degree lower than $4$.
\end{example}

\begin{example}
    In the continuum limit of the Toda lattice \cite{DM}, an external
    field of the form
    \[
        V_{t_1,t_2}(x)=(1+t_1)(V_0(x)+t_2x)
    \]
    has been studied. This deformation of $V_0$ can be written in the form
    (\ref{Vst}) (so that it is included in the class of external fields studied in this paper).
    Indeed, if we let $V_1(x)=x$ and $V_2(x)=V_0(x)+cx$,
    with $c$ some constant chosen such that the critical condition (\ref{conditionV2})
    holds, then
    \[V_{t_1,t_2}=V_0+sV_1+tV_2,\] with $s=t_2+t_1t_2-ct_1$ and
    $t=t_1$.
\end{example}

\begin{remark}
    In Section \ref{section: equilibrium measures} we will show that
    assumption (iii) is equivalent to the vanishing of the
    equilibrium density $\frac{d\nu_2(x)}{dx}$ at the right endpoint $b$,
    where $\nu_2$ is the unique measure which minimizes $I_{V_2}(\nu)$,
    see (\ref{definition: energy}), among all signed measures $\nu$,
    supported on $[a,b]$ and having zero mass, $\nu([a,b])=0$.
\end{remark}

\begin{remark}
    The case where the left (instead of the right) endpoint of the
    support is singular can be transformed to our case by considering
    the external field $V_{s,t}(-x)$.
\end{remark}

\begin{remark}
    Without giving any mathematical details, we now describe the transitions
    that can occur for $s$ and $t$ near 0.
    First, if we let $t=0$ and $s$ vary around $0$, one typically
    observes the transition from the regular one-interval case to the
    singular case and back to the regular one-interval case. Next, for $s=0$
    and $t$ around $0$, we can observe the transition from
    the regular one-interval case to the regular two-interval case.
    Finally, letting both $s$ and $t$ vary around 0, we can observe one of the above described
    transitions, or the critical transition where a type II singular
    point moves to the endpoint $b$, where it becomes a type III
    singular point before moving on as a type I singular point.
\end{remark}

Further, to describe our results, we have to introduce constants
$c,c_1$, and $c_2$,
\begin{equation}\label{definition: cc1c2}
    c=\left(\frac{15}{2}h_0''(b)\sqrt{b-a}\right)^{2/7}> 0,
    \qquad c_1=\frac{h_1(b)}{c^{1/2}(b-a)^{1/2}},
    \qquad c_2=-\frac{h_2'(b)}{c^{3/2}(b-a)^{1/2}},
\end{equation}
where $h_0$ is the real analytic function appearing in
(\ref{psi0h0}), and where the functions $h_1$ and $h_2$ are defined
as,
\begin{equation}
    h_j(x)=-\frac{1}{\pi}\PVint_a^b\sqrt{(b-u)(u-a)}V_j'(u)\frac{du}{u-x},
        \qquad\mbox{for $x\in[a,b]$ and $j=1,2$.}
\end{equation}

\subsubsection{Universality of the double scaling limit}

Our main result is the following.

\begin{theorem}\label{theorem: universality}
    Let $V_{s,t}=V_0+sV_1+tV_2$ be such that Assumptions {\rm \ref{assumptions}}
    above are satisfied. We take a double scaling limit where we let
    $n\to\infty$ and at the same time $s,t\to 0$, in such a way that
    $\lim n^{6/7}s$ and $\lim n^{4/7}t$ exists, and put
    \begin{equation}\label{lim stn-1}
        s_0=c_1\cdot\,\lim n^{6/7}s\in\mathbb R,
            \qquad t_0=c_2\cdot\,\lim n^{4/7}t\in\mathbb R,
    \end{equation}
    where the constants $c_1$ and $c_2$ are defined by
    {\rm (\ref{definition: cc1c2})}. Then, the 2-point kernel $K_n^{(s,t)}$ satisfies the
    following universality result,
    \begin{equation}
        \lim
        \frac{1}{cn^{2/7}}K_n^{(s,t)}\left(b+\frac{u}{cn^{2/7}},b+\frac{v}{cn^{2/7}}\right)
        =K^{{\rm crit, III}}(u,v;s_0,t_0),
    \end{equation}
    uniformly for $u,v$ in compact subsets of $\mathbb R$. Here,
    $K^{{\rm crit, III}}$ is built out of the functions $\Phi_1$
    and $\Phi_2$ defined in Section {\rm \ref{subsection: PI2 equation}},
    \begin{equation}\label{definition: Kcrit}
        K^{{\rm crit, III}}(u,v;s,t)=\frac{\Phi_1(u;s,t)\Phi_2(v;s,t)-\Phi_1(v;s,t)\Phi_2(u;s,t)}{2\pi
        i(u-v)}.
    \end{equation}
\end{theorem}

\begin{remark}
    Since $y(s,t)$ has no poles \cite{CV} for $s,t\in\mathbb R$, the kernel
    $K^{{\rm crit, III}}(u,v;s,t)$ exists for all real $u, v, s$, and
    $t$. Furthermore, using a similar argument as in \cite[Lemma 2.3
    (ii)]{CV}, one can show that $e^{\pi i/4}\Phi_1$ and $e^{\pi
    i/4}\Phi_2$ are real. It then follows that $K^{{\rm crit,
    III}}(u,v;s,t)$ is real for real $u$, $v$, $s$, and $t$.
\end{remark}

\begin{remark}
    It is possible to give an integral formula for $K^{{\rm crit, III}}$.
    Using the fact that
    $\bigl(\begin{smallmatrix}\Phi_1 \\ \Phi_2\end{smallmatrix}\bigr)$
    satisfies the second differential equation of the Lax pair
    (\ref{differential equations}), we have that
    \[
        \frac{\partial \Phi_1}{\partial s}(\zeta;s,t)=\Phi_2(\zeta;s,t),
        \qquad\mbox{and}\qquad
        \frac{\partial \Phi_2}{\partial s}(\zeta;s,t)=(\zeta -
        2y(s,t))\Phi_1(\zeta;s,t).
    \]
    Using (\ref{definition: Kcrit}) this yields,
    \[
        \frac{\partial K^{{\rm crit, III}}}{\partial s}(u,v;s,t)
            = -\frac{1}{2\pi i} \Phi_1(u;s,t)\Phi_1(v;s,t).
    \]
    Now, since $\lim_{s\to -\infty}K^{{\rm crit, III}}(u,v;s,t)=0$, which can
    be shown using a Deift/Zhou steepest descent method argument
    \cite{DeiftZhou}, it then follows that $K^{{\rm crit, III}}$ has
    the following integral formula,
    \begin{equation}
        K^{{\rm crit, III}}(u,v;s,t)
            = -\frac{1}{2\pi i}\int_{-\infty}^s
            \Phi_1(u;\sigma,t)\Phi_1(v;\sigma,t)d\sigma.
    \end{equation}
\end{remark}

\begin{remark}
    Theorem \ref{theorem: universality} can be generalized to the case
    where the support of $\nu_0$ (the equilibrium measure in external field $V_0$)
    consists of more than one interval.
    Then, the proof becomes much more technical,
    although the main ideas remain the same. We comment in Remark
    \ref{remark: multiple intervals} on the modifications that have to
    be made in the multi-interval case.
\end{remark}

\subsubsection{Recurrence coefficients for orthogonal polynomials}

It is well-known \cite{Szego} that the orthonormal polynomials
$p_k=p_k^{(n,s,t)}$ satisfy a three-term recurrence relation of the
form,
\begin{equation}\label{three-term recurrence relation}
    xp_k(x)=a_{k+1}p_{k+1}(x)+b_k p_k(x)+a_k p_{k-1}(x),
\end{equation}
where $a_k=a_k^{(n,s,t)}>0$ and $b_k=b_k^{(n,s,t)}\in\mathbb{R}$ (we suppress
the $s$ and $t$ dependence for brevity). In the generic case where $V_0$ has no
singular points, the recurrence coefficients for $s=t=0$ have the following
asymptotics, see e.g.\ \cite{BI2, Deift},
\begin{equation}\label{recurrence-generic}
    a_n^{(n,0,0)}=\frac{b-a}{4}+\bigO(n^{-1}),
        \qquad b_n^{(n,0,0)}=\frac{b+a}{2}+\bigO(n^{-1}),
        \qquad\mbox{as $n\to\infty$.}
\end{equation}
For singular potentials $V_0$, the constant terms in the expansions
(\ref{recurrence-generic}) remain the same, but the error terms
behave differently \cite{BI2, CKV}. In our case of interest, where
we have a type III singular (edge) point of $V_0$ with $m=1$, the
error term is of order $\bigO(n^{-2/7})$, and the coefficient of the
$n^{-2/7}$ term is expressed in terms of the special solution $y$ of
the $P_I^2$ equation discussed in Section \ref{subsection: PI2
equation}.

\begin{theorem}\label{theorem: recurrence coefficients}
    Let $V_{s,t}$ be such that Assumptions {\rm \ref{assumptions}}
    above are satisfied. Consider the three-term recurrence relation
    {\rm (\ref{three-term recurrence relation})} satisfied by the
    orthonormal polynomials $p_k=p_k^{(n,s,t)}$ with respect
    to the weight function $e^{-nV_{s,t}}$. Then, in the double
    scaling limit where $n\to\infty$ and $s,t\to 0$, in such a way that
    $\lim n^{6/7}s$ and $\lim n^{4/7}t$ exists, and put
    \begin{equation}
        s_0=c_1\cdot\,\lim n^{6/7}s\in\mathbb R,
            \qquad t_0=c_2\cdot\,\lim n^{4/7}t\in\mathbb R,
    \end{equation}
    with $c_1$ and $c_2$ given by {\rm (\ref{definition: cc1c2})},
    we have
    \begin{align}\label{theorem: recurrence coefficients: anst}
        a_n^{(n,s,t)} &=\frac{b-a}{4}+\frac{1}{2c}\, y(c_1n^{6/7}s,c_2n^{4/7}t)n^{-2/7}+\bigO(n^{-3/7}),
        \\[2ex]
        \nonumber
        & = \frac{b-a}{4}+\frac{1}{2c}\, y(s_0,t_0)n^{-2/7}(1+o(1)),
    \end{align}
    and
    \begin{align}\label{theorem: recurrence coefficients: bnst}
        b_n^{(n,s,t)} &=\frac{b+a}{2}+\frac{1}{c}\, y(c_1n^{6/7}s,c_2n^{4/7}t)n^{-2/7}+\bigO(n^{-3/7})
        \\[2ex]
        \nonumber
        & = \frac{b+a}{2}+\frac{1}{c}\, y(s_0,t_0)n^{-2/7}(1+o(1)),
    \end{align}
    where the constant $c$ is given by {\rm (\ref{definition:
    cc1c2})}, and where $y$ is the special solution of the $P_I^2$ equation discussed
    in Section {\rm \ref{subsection: PI2 equation}}.
\end{theorem}

\begin{remark}
    Note that the expansions of the recurrence coefficients are
    of the same form as the conjectured (by Dubrovin \cite[Main Conjecture, Part 3]{Dubrovin}, see also
    \cite{DubrovinI}) expansions for solutions of
    perturbed hyperbolic equations. Here, the perturbation parameter
    $\epsilon$ plays the role of $1/n$ in our context.
\end{remark}

\begin{remark}
    For polynomials which are orthogonal on certain complex contours, it
    can occur that the equilibrium density vanishes like a power $3/2$.
    Asymptotics of the recurrence coefficients in this case were
    obtained in \cite{DK}. Here, a special solution of the Painlev\'e I equation occurs instead of
    a solution of the $P_I^2$ equation and the asymptotics are in powers of
    $n^{-1/5}$.

    Observe further that in \cite{DK} there is no term
    of order $n^{-1/5}$ in the asymptotics. In (\ref{theorem: recurrence coefficients: anst}) and
    (\ref{theorem: recurrence coefficients: bnst}) we see that there is no term of order $n^{-1/7}$.
    In the proof of Theorem \ref{theorem: recurrence coefficients} this term will drop out in a similar
    way as the $n^{-1/5}$-term in \cite{DK}.
\end{remark}

\subsection{Outline of the rest of the paper}

We prove our results by characterizing the orthogonal polynomials via the
well-known $2\times 2$ matrix valued Fokas-Its-Kitaev Riemann-Hilbert (RH)
problem \cite{FokasItsKitaev} and applying the Deift/Zhou steepest descent
method \cite{DeiftZhou} to analyze this RH problem asymptotically. This
approach has been used many times before, see e.g.\ \cite{CK, CKV, Deift,
DKMVZ2, DKMVZ1, DK, KMVV, Vanlessen, Vanlessen2}.

An important step in the Deift/Zhou steepest descent method is the
construction of so-called $g$-functions associated with equilibrium
measures. Those equilibrium measures will be constructed in Section
\ref{section: equilibrium measures}. In order to deal with the
deformations $V_{s,t}$ of $V_0$, we use modified equilibrium
problems where we allow the measures to be negative, which was also
done in \cite{CK, CKV, DK}. Another modification of the equilibrium
problem is that we choose the support of the equilibrium measure
fixed, instead of allowing it to choose its own support.

In Section \ref{section: sd}, we perform the Deift/Zhou steepest
descent analysis to the RH problem $Y$ for orthogonal polynomials.
Via a series of transformations $Y\mapsto T\mapsto S\mapsto R$ we
want to arrive at a RH problem for $R$ which is normalized at
infinity (i.e.\ $R(z)\to I$ as $z\to\infty$) and with jumps
uniformly close to the identity matrix. Then, $R$ itself is close to
the identity matrix. By unfolding the series of transformations we
then get the asymptotics of $Y$. The key step in this method will be
the local analysis near the endpoints $a$ and $b$. Near the regular
endpoint $a$, we construct (in Section \ref{subsection: parametrix
near a}) a parametrix built out of Airy functions. Due to the
modified equilibrium measures, which have a fixed support, we also
need to make a technical modification in the construction of the
Airy parametrix, compared with the parametrix as used e.g.\ in
\cite{Deift}. To construct the local parametrix near the singular
endpoint $b$ (in Section \ref{subsection: parametrix near b}) we use
a model RH problem associated with the special solution $y$ of the
$P_I^2$ equation as discussed in Section \ref{subsection: PI2
equation}.

The results of Section \ref{section: sd} will be used in Section
\ref{section: universality} to prove the universality result for the
correlation kernel (see Theorem \ref{theorem: universality}) and in
Section \ref{section: recurrence} to determine the asymptotics of
the recurrence coefficients (see Theorem \ref{theorem: recurrence
coefficients}).

\section{Equilibrium measures}\label{section: equilibrium measures}

We consider external fields $V_{s,t}=V_0+sV_1+tV_2$ which satisfy
Assumptions \ref{assumptions} in the beginning of Section
\ref{subsection: statement of results}. In order to perform the
Deift/Zhou steepest descent analysis to the RH problem for
orthogonal polynomials one would expect to use the equilibrium
measure $\mu_{s,t}$ in external field $V_{s,t}$ minimizing
$I_{V_{s,t}}(\mu)$, see (\ref{definition: energy}), among all
probability measures $\mu$ on $\mathbb R$. However, as in
\cite{CK,CKV,DK} it will be more convenient to use modified
equilibrium measures $\nu_{s,t}$ which we allow to be negative.
Furthermore, unlike in \cite{CK,CKV,DK}, we take the support of the
measures $\nu_{s,t}$ to be fixed instead of letting it depend on $s$
and $t$.

The aim of this section is to find measures $\nu_{s,t}$ (depending
on the parameters $s,t\in\mathbb R$) supported on the interval
$[a,b]\subset \mathbb R$ (where $[a,b]$ is the support of the
equilibrium measure $\nu_0$ in external field $V_0$), such that
$\nu_{s,t}([a,b])=1$, and such that they satisfy the following
condition: there exist $\ell_{s,t}\in\mathbb R$ such that for
every $\delta>0$ there are $\varepsilon,\kappa>0$ sufficiently
small such that for $s,t\in[-\varepsilon,\varepsilon]$,
\begin{align}
    \label{nust: variational equality}
    & 2\int\log|x-u|d\nu_{s,t}(u)-V_{s,t}(x)=\ell_{s,t},
        &&\mbox{for $x\in[a,b]$.}
    \\
    \label{nust: variational inequality}
    & 2\int\log|x-u|d\nu_{s,t}(u)-V_{s,t}(x)<\ell_{s,t}-\kappa,
        &&\mbox{for $x\in\mathbb R\setminus [a-\delta, b+\delta]$.}
\end{align}

\medskip

We seek $\nu_{s,t}$ in the following form,
\begin{equation}\label{definition: nust}
    \nu_{s,t}=\nu_0+s\nu_1+t\nu_2,
\end{equation}
where $\nu_0$ is the equilibrium measure in external field $V_0$ minimizing
$I_{V_0}(\nu)$, see (\ref{definition: energy}), among all probability measures
$\nu$ on $\mathbb R$. From Assumption \ref{assumptions} (ii) we know that
$\nu_0$ can be written as follows
\begin{equation}\label{definition: h0}
    d\nu_0(x)=\psi_{0,+}(x)\chi_{[a,b]}(x)dx,
\end{equation}
where $\chi_{[a,b]}$ is the indicator function of the set $[a,b]$,
and where $\psi_{0,+}$ is the $+$boundary value of the function
\begin{equation}\label{definition: psi0}
    \psi_0(z)=\frac{1}{2\pi
        i}R(z) h_0(z),\qquad\mbox{for $z\in\mathcal V\setminus[a,b]$,}
\end{equation}
with $h_0$ analytic in the neighborhood $\mathcal V$ of the real
line, and with
\begin{equation}\label{definition: R}
    R(z)=\bigl((z-a)(z-b)\bigr)^{1/2},
        \qquad\mbox{for $z\in\mathbb{C}\setminus[a,b]$.}
\end{equation}
Here, we take the principal branch of the square root so that $R$ is
analytic in $\mathbb{C}\setminus[a,b]$. Further, since $a$ is a
regular (edge) point and since $b$ is a type III singular (edge)
point with $m=1$, we have, cf.\ (\ref{assumptions: eq1}) and
(\ref{assumptions: eq2}),
\begin{equation}\label{h0(a)=0 etc}
    h_0(a)>0,\qquad
    h_0(b)=h_0'(b)=0,\qquad\mbox{and}\qquad h_0''(b)>0.
\end{equation}
Since $V_0$ is assumed to have no other singular points besides $b$,
we know (cf.\ (\ref{variationalcondition:must-equality1}) and
(\ref{variationalcondition:must-inequality1})) that $\nu_0$
satisfies the following condition: there exists
$\ell_0\in\mathbb{R}$ such that
\begin{align}
    \label{variationalcondition:nu0-equality}
    & 2\int\log|x-u|d\nu_0(u)-V_0(x)=\ell_0, && \mbox{for $x\in [a,b]$,} \\
    \label{variationalcondition:nu0-inequality}
    & 2\int\log|x-u|d\nu_0(u)-V_0(x)< \ell_0, && \mbox{for $x\in \mathbb R\setminus [a,b]$.}
\end{align}

\medskip

We will now construct the two measures $\nu_1$ and $\nu_2$. In order
to do this we introduce the following auxiliary (analytic)
functions,
\begin{equation}\label{definition: hj}
    h_j(z)=\frac{1}{2\pi i}\oint_\gamma
    R(\xi)V_j'(\xi)\frac{d\xi}{\xi-z},\qquad \mbox{for $z\in\mathcal V$ and
    $j=1,2$,}
\end{equation}
where $\gamma$ is a positively oriented contour in $\mathcal V$ with
$[a,b]$ and $z$ in its interior, and where $R$ is given by
(\ref{definition: R}). Observe that, using the fractional residue
theorem, one has,
\begin{equation}\label{hj real}
    h_j(x)=-\frac{1}{\pi i}\PVint_a^b
    R_+(u)V_j'(u)\frac{du}{u-x},\qquad \mbox{for $x\in[a,b]$,}
\end{equation}
where the integral is a Cauchy principal value integral. So, $h_j$ is real on
$[a,b]$. Observe that by Assumption \ref{assumptions} (iii) and (\ref{hj
real}),
\begin{equation}\label{h2(b)=0}
    h_2(b)=0.
\end{equation}

\begin{lemma}
    Define two signed measures
    $\nu_1$ and $\nu_2$ supported on $[a,b]$ as
    \begin{equation}
        d\nu_j(x)=\psi_{j,+}(x)\chi_{[a,b]}dx,\qquad j=1,2,
    \end{equation}
    where $\chi_{[a,b]}$ is the indicator function of the
    set $[a,b]$, and where $\psi_{j,+}$ is the $+$boundary value of
    the function
    \begin{equation}\label{definition: psij}
        \psi_j(z)=\frac{1}{2\pi
        i}\frac{h_j(z)}{R(z)},\qquad\mbox{for $z\in\mathcal V\setminus[a,b]$.}
    \end{equation}
    Here, $h_j$ is given by {\rm(\ref{definition: hj})}, see also {\rm (\ref{hj real})} for its expression on $[a,b]$, and
    $R$ is given by {\rm (\ref{definition: R})}. Then, $\nu_j$ has zero
    mass, i.e.
    \begin{equation}\label{zero mass nuj}
        \nu_j([a,b])=\int_a^b\psi_{j,+}(u)du=0,
    \end{equation}
    and there exist constants $\ell_j\in\mathbb R$ such that
    \begin{equation}\label{variational equality: nuj}
        2\int\log|x-u|d\nu_j(u)-V_j(x)=\ell_j, \qquad\mbox{for $x\in[a,b]$.}
    \end{equation}
\end{lemma}

\begin{proof}
    Define, for $j=1,2$, the auxiliary functions
    \begin{equation}\label{definition: F}
        F_j(z)=\frac{1}{2\pi iR(z)}\int_a^b
        R_+(u)V_j'(u)\frac{du}{u-z},\qquad\mbox{for $z\in\mathbb
        C\setminus[a,b]$,}
    \end{equation}
    which, by standard techniques and by (\ref{definition: hj}) and (\ref{definition: psij}),
    are equal to
    \begin{align*}
        F_j(z)&=\frac{1}{2}V_j'(z)-\frac{1}{4\pi i R(z)}\oint_\gamma
        R(\xi)V_j'(\xi)\frac{d\xi}{\xi-z}\\[1ex]
        &=\frac{1}{2}V_j'(z)-\pi i\psi_j(z),&\mbox{for $z\in\mathcal V\setminus[a,b]$,}
    \end{align*}
    where $\gamma$ is a positively oriented contour in $\mathcal V$ with
    $[a,b]$ and $z$ in its interior.
    This, together with the fact that $\psi_{j,+}=-\psi_{j,-}$ on $(a,b)$, yields
    \begin{align}
        \label{SP1}
        & F_{j,+}(x)-F_{j,-}(x)=-2\pi i\psi_{j,+}(x), &\mbox{for $x\in[a,b]$,}
        \\[1ex]
        \label{SP2}
        &F_{j,+}(x)+F_{j,-}(x)=V_j'(x), &\mbox{for $x\in[a,b]$.}
    \end{align}

    Since $F_j$ is analytic in $\mathbb
    C\setminus[a,b]$ and since, by (\ref{definition: F}), $F_j(z)=\bigO(z^{-2})$ as
    $z\to\infty$, a standard complex analysis argument, shows that
    \[
        \frac{1}{2\pi
        i}\int_a^b\frac{F_{j,+}(u)-F_{j,-}(u)}{u-z}ds=F_j(z),\qquad\mbox{for $z\in\mathbb C\setminus[a,b]$.}
    \]
    By (\ref{SP1}), this yields,
    \[
        F_j(z)=-\int_a^b\frac{\psi_{j,+}(u)}{u-z}du=
        -z^{-1}\int_a^b\psi_{j,+}(u)du+\bigO(z^{-2}),\qquad\mbox{as $z\to\infty$.}
    \]
    Comparing this with the fact that
    $F_j(z)=\bigO(z^{-2})$ as
    $z\to\infty$, we obtain $\int_a^b\psi_{j,+}(u)du=0$, so that (\ref{zero mass nuj}) is proven.

    It remains to prove (\ref{variational equality: nuj}).
    It is straightforward to check that,
    \[
        F_j(z)=-\int_a^b\frac{\psi_{j,+}(u)}{u-z}du=-\pi
        i\psi_j(z)+\frac{1}{2}\oint_\gamma\frac{\psi_j(\xi)}{\xi-z}d\xi,
        \qquad\mbox{for $z\in\mathcal V\setminus[a,b]$,}
    \]
    so that, using the fractional residue theorem,
    \[
        F_{j,\pm}(x)=-\pi
        i\psi_{j,\pm}(x)-\PVint_a^b\frac{\psi_{j,+}(u)}{u-x}du,\qquad\mbox{for $x\in[a,b]$.}
    \]
    From (\ref{SP2}) and the fact that $\psi_{j,+}+\psi_{j,-}=0$ on $[a,b]$ this yields,
    \begin{equation}
        \frac{d}{dx}\left(2\int\log|x-u|d\nu_j(u)+V_j(x)\right)=
        2\PVint_a^b\frac{\psi_{j,+}(u)}{u-x}du+\bigl(F_{j,+}(x)+F_{j,-}(x)\bigr)=0.
    \end{equation}
    This proves (\ref{variational equality: nuj}).
\end{proof}

\begin{corollary}\label{corollary: equilibrium measures}
    Let $\nu_{s,t}=\nu_0+s\nu_1+t\nu_t$. Then,
    $d\nu_{s,t}(x)=\psi_{s,t,+}(x)\chi_{[a,b]}dx$, where
    \begin{equation}\label{definition: psist}
        \psi_{s,t}=\psi_0+s\psi_1+t\psi_2,\qquad\mbox{on $\mathcal
        V\setminus[a,b]$,}
    \end{equation}
    with $\psi_0$ given by {\rm (\ref{definition: psi0})} and
    $\psi_{1}$ and $\psi_2$ given by {\rm (\ref{definition: psij})}.
    So, $\nu_{s,t}$ is supported on $[a,b]$ and has mass one, i.e.\ $\nu_{s,t}([a,b])=1$.
    Further, there exist constants $\ell_{s,t}\in\mathbb{R}$ such
    that for any $\delta>0$ there are $\varepsilon,\kappa>0$
    sufficiently small such that for
    $s,t\in[-\varepsilon,\varepsilon]$ the conditions {\rm (\ref{nust: variational equality})} and
    {\rm (\ref{nust: variational inequality})} are satisfied.
\end{corollary}

\begin{proof}
    Since $\nu_{s,t}=\nu_0+s\nu_1+t\nu_t$, from (\ref{zero mass
    nuj}), and from the fact that $\nu_0([a,b])=1$ it is clear that $\nu_{s,t}([a,b])=1$. Next, with
    $\ell_{s,t}=\ell_0+s\ell_1+t\ell_2$, we have
    \begin{equation}\label{proof: corrolary equilibrium measures: eq1}
        2\int \log|x-u|d\nu_{s,t}(u)-V_{s,t}(x)-\ell_{s,t}= I_0(x)+s
        I_1(x)+t I_2(x)
    \end{equation}
    where
    \[
        I_j(x)= 2\int
        \log|x-u|d\nu_j(u)-V_j(x)-\ell_j,\qquad j=1,2,3.
    \]
    Then, condition (\ref{nust: variational equality}) follows from
    (\ref{variationalcondition:nu0-equality}) and (\ref{variational equality:
    nuj}). Now, by using (\ref{variationalcondition:nu0-inequality}) and the fact that $I_0(x)\to-\infty$ as
    $|x|\to\infty$, there exists $\kappa>0$ such that
    \begin{equation}\label{proof: corrolary equilibrium measures: eq2}
        I_0<-\frac{3}{2}\kappa,\qquad\mbox{on
        $\mathbb{R}\setminus[a-\delta,b+\delta]$.}
    \end{equation}
    Further, one can check that $I_1$ and $I_2$ are bounded on
    $\mathbb{R}\setminus[a-\delta,b+\delta]$, and thus there exists
    $\varepsilon>0$ such that for
    $s,t\in[-\varepsilon,\varepsilon]$,
    \begin{equation}\label{proof: corrolary equilibrium measures: eq3}
        s I_1+t I_2<\frac{1}{2}\kappa,\qquad\mbox{on
        $\mathbb{R}\setminus[a-\delta,b+\delta]$.}
    \end{equation}
    Inserting (\ref{proof: corrolary equilibrium measures: eq2}) and
    (\ref{proof: corrolary equilibrium measures: eq3}) into
    (\ref{proof: corrolary equilibrium measures: eq1}) we obtain condition
    (\ref{nust: variational inequality}).
\end{proof}

\begin{remark}\label{remark: nust positive}
The measure $\nu_1$ ($\nu_2$) is the equilibrium measure that minimizes
$I_{V_1}(\nu)$ ($I_{V_2}(\nu)$) among all signed measures $\nu$, supported on
$[a,b]$ with $\nu([a,b])=0$. The measures $\nu_{s,t}$ on the other hand
minimize $I_{V_{s,t}}(\nu)$ among all signed measures supported on $[a,b]$ with
$\nu([a,b])=1$.

Observe that since $\nu_0$ has a strictly positive density on
$(a,b)$ (since $\nu_0$ has no type II singular points) we have for
any $\delta>0$ that $\nu_{s,t}$ is positive on $(a+\delta,b-\delta)$
for $s,t$ sufficiently small.
\end{remark}

\section{Riemann-Hilbert analysis}
    \label{section: sd}

\subsection{RH problem for orthogonal polynomials}

For each fixed $n, s,$ and $t$, we consider the Fokas-Its-Kitaev
Riemann-Hilbert problem \cite{FokasItsKitaev} characterizing the
orthogonal polynomials $p_k^{(n,s,t)}$ with respect to the weight
functions $e^{-nV_{s,t}}$. We seek a $2\times 2$ matrix-valued
function $Y(z)=Y(z;n,s,t)$ (we suppress the $n,s,$ and $t$
dependence for brevity) that satisfies the following conditions.

\subsubsection*{RH problem for $Y$:}

\begin{itemize}
    \item[(a)] $Y:\mathbb{C}\setminus \mathbb{R}\to\mathbb{C}^{2\times 2}$ is analytic.
    \item[(b)] $Y$ possesses continuous boundary values for $x\in\mathbb{R}$ denoted by
        $Y_+(x)$ and $Y_-(x)$, where $Y_+(x)$ and $Y_-(x)$ denote the limiting values of
        $Y(z')$ as $z'$ approaches $x$ from above and below, respectively, and
        \begin{equation}\label{RHP Y: b}
            Y_+(x)=Y_-(x)
                \begin{pmatrix}
                    1 & e^{-nV_{s,t}(x)} \\
                    0 & 1
                \end{pmatrix},
                \qquad  \mbox{for $x \in\mathbb{R}$.}
        \end{equation}
    \item[(c)] $Y$ has the following asymptotic behavior at infinity
        \begin{equation}\label{RHP Y: c}
            Y(z)=\left(I+\bigO(z^{-1})\right)
                \begin{pmatrix}
                    z^n & 0 \\
                    0 & z^{-n}
                \end{pmatrix},
                \qquad  \mbox{as $z\rightarrow \infty$.}
        \end{equation}
\end{itemize}

\medskip

The unique solution of the RH problem is given by
\begin{equation}\label{RHP Y: solution}
    Y(z)=
    \begin{pmatrix}
        \kappa_n^{-1}p_n(z) &
            \displaystyle{\frac{\kappa_n^{-1}}{2\pi i}\int_{\mathbb R}\frac{p_n(u) e^{- n V_{s,t}(u)}}{u-z}\,
            du}
        \\[3ex]
        - 2\pi i \kappa_{n-1} p_{n-1}(z) &
            \displaystyle{-\kappa_{n-1} \int_{\mathbb R}\frac{p_{n-1}(u) e^{- n
            V_{s,t}(u)}}{u-z}\,du}
    \end{pmatrix},\qquad\mbox{for $z\in\mathbb C \setminus\mathbb R$,}
\end{equation}
where $p_k=p_k^{(n,s,t)}$ is the $k$-th degree orthonormal
polynomial with respect to the varying weight $e^{-nV_{s,t}}$, and
where $\kappa_k=\kappa_k^{(n,s,t)}>0$ is the leading coefficient of
$p_k$. The solution (\ref{RHP Y: solution}) is due to Fokas, Its,
and Kitaev \cite{FokasItsKitaev}, see also
\cite{Deift,DKMVZ2,DKMVZ1}.

It is now possible to write the 2-point kernel $K_n^{(s,t)}$, see
(\ref{kernel}), in terms of $Y$. Indeed using the
Christoffel-Darboux formula for orthogonal polynomials and the fact
that $\det Y \equiv 1$ (which follows easily from (\ref{RHP Y: b}),
(\ref{RHP Y: c}), and Liouville's theorem), we get
\begin{equation}\label{KinY}
    K_n^{(s,t)}(x,y) =
        e^{-\frac{n}{2}V_{s,t}(x)}e^{-\frac{n}{2}V_{s,t}(y)} \frac{1}{2\pi i(x-y)}
        \begin{pmatrix}
            0 & 1
        \end{pmatrix}
        Y_{\pm}^{-1}(y) Y_{\pm}(x)
        \begin{pmatrix}
            1 \\ 0
        \end{pmatrix}.
\end{equation}
So, in order to prove Theorem \ref{theorem: universality}, we need
to analyze the RH problem for $Y$ asymptotically. We do this by
applying the Deift/Zhou steepest descent method \cite {DeiftZhou} to
this RH problem.

\subsection{Normalization of the RH problem at infinity: $Y\mapsto T$}
    \label{subsection: sd-normalization}

In order to normalize the RH problem for $Y$ at infinity, the
equilibrium measures $\nu_{s,t}$, introduced in Section
\ref{section: equilibrium measures} play a key role. Consider the
$\log$-transform $g_{s,t}$ of $\nu_{s,t}$,
\begin{equation}\label{definition: gst}
    g_{s,t}(z)=\int_a^b\log(z-u)d\nu_{s,t}(u),
        \qquad\mbox{for $z\in\mathbb C\setminus(-\infty,b]$.}
\end{equation}
Here, we take the principal branch of the logarithm so that
$g_{s,t}$ is analytic in $\mathbb C \setminus (-\infty,b]$. We now
give properties of $g_{s,t}$ which are crucial in the following.
From (\ref{definition: gst}) and condition (\ref{nust: variational
equality}) it follows that
\begin{equation}\label{property g: 1}
    g_{s,t,+}(x)+g_{s,t,-}(x)-V_{s,t}(x)-\ell_{s,t}=0,\qquad\mbox{for
    $x\in[a,b]$.}
\end{equation}
Another crucial property is that
\begin{equation}\label{property g: 2}
    g_{s,t,+}(x)-g_{s,t,-}(x)=2\pi i\int_x^b d\nu_{s,t}(u),\qquad \mbox{for $x\in\mathbb
    R$,}
\end{equation}
so that since $\nu_{s,t}$ is supported on $[a,b]$ and has mass one
(see Corollary \ref{corollary: equilibrium measures}),
\begin{equation}\label{property g: 3}
    g_{s,t,+}(x)-g_{s,t,-}(x)=
    \begin{cases}
        2\pi i, & \mbox{for $x<a$,} \\
        0, & \mbox{for $x>b$.}
    \end{cases}
\end{equation}

\medskip

Now, we are ready to perform the first transformation $Y\mapsto T$.
Define the matrix valued function $T$ as
\begin{equation}\label{TinY}
    T(z)=e^{-\frac{1}{2} n\ell_{s,t}\sigma_3} Y(z) e^{-ng_{s,t}(z)\sigma_3}
    e^{\frac{1}{2}n\ell_{s,t}\sigma_3}, \qquad \mbox{for
    $z\in\mathbb{C}\setminus\mathbb R$,}
\end{equation}
where $\ell_{s,t}$ is the constant that appears in the variational
conditions (\ref{nust: variational equality}) and (\ref{nust:
variational inequality}), and where $\sigma_3 =
\left(\begin{smallmatrix} 1 & 0
\\ 0 & -1
\end{smallmatrix}\right)$ denotes the third Pauli-matrix.
Using (\ref{property g: 1}), (\ref{property g: 3}), the RH
conditions for $Y$, and the fact that $g_{s,t}(z)=\log z+\bigO(1/z)$
as $z\to\infty$, it is straightforward to check that $T$ is a
solution to the following RH problem.

\subsubsection*{RH problem for $T$:}
\begin{itemize}
    \item[(a)] $T:\mathbb{C}\setminus \mathbb{R}\to\mathbb{C}^{2\times 2}$ is analytic.
    \item[(b)] $T_+(x)=T_-(x)v_T(x)$ for $x\in\mathbb{R}$, with
        \begin{equation} \label{RHP T: b}
            v_T=
            \begin{cases}
                \begin{pmatrix}
                    e^{-n(g_{s,t,+}-g_{s,t,-})} & 1 \\
                    0 & e^{n(g_{s,t,+}-g_{s,t,-})}
                \end{pmatrix}, & \mbox{on $(a,b)$,} \\[3ex]
                \begin{pmatrix}
                    1 & e^{n(g_{s,t,+}+g_{s,t,-}-V_{s,t}- \ell_{s,t})} \\
                    0 & 1
                \end{pmatrix}, & \mbox{on $\mathbb R\setminus (a,b)$.}
            \end{cases}
        \end{equation}
    \item[(c)] $T(z)=I+\bigO(1/z)$,\qquad as $z\to\infty$.
\end{itemize}

\begin{remark}
    From (\ref{property g: 2}) we see that the diagonal entries of
    $v_T$ on $(a,b)$ are rapidly oscillating for large $n$.
    Further, using condition (\ref{nust: variational inequality}) and
    (\ref{definition: gst}), we see that $v_T-I$ decays
    exponentially on $\mathbb R\setminus[a-\delta,b+\delta]$.
\end{remark}

\subsection{Opening of the lens: $T\mapsto S$}
    \label{subsection: sd-lens}

Here, we will transform the oscillatory diagonal entries of the jump
matrix $v_T$ on $(a,b)$ into exponentially decaying off-diagonal
entries. This step is referred to as the opening of the lens.

\medskip

Introduce a scalar function $\phi_{s,t}$ as,
\begin{equation}\label{definition: phist}
    \phi_{s,t}(z) =-\pi i\int_z^b\psi_{s,t}(\xi)d\xi,\qquad\mbox{for
    $z\in\mathcal V\setminus(-\infty,b]$,}
\end{equation}
where the path of integration does not cross the real line, and
where $\psi_{s,t}$ is defined by (\ref{definition: psist}).

The important feature of the function $\phi_{s,t}$ is that by
(\ref{property g: 2}), $\phi_{s,t,+}$ and $\phi_{s,t,-}$ are purely
imaginary on $(a,b)$ and satisfy,
\begin{equation}
    -2\phi_{s,t,+}(x)=2\phi_{s,t,-}(x)=2\pi i\int_x^b d\nu_{s,t}(u)=g_{s,t,+}(x)-g_{s,t,-}(x),
    \quad\mbox{for $x\in(a,b)$,}
   \label{equation-phig-bulk}
\end{equation}
which means that $-2\phi_{s,t}$ and $2\phi_{s,t}$ provide analytic
extensions of $g_{s,t,+}-g_{s,t,-}$ into the upper half-plane and
lower half-plane, respectively. Further,
$2g_{s,t}+2\phi_{s,t}-V_{s,t}-\ell_{s,t}$ is analytic in
$\mathcal{V}\setminus(-\infty,b]$ and satisfies by
(\ref{equation-phig-bulk}) and (\ref{property g: 1}),
\[
    2g_{s,t,\pm}+2\phi_{s,t,\pm}-V_{s,t}-\ell_{s,t}=g_{s,t,+}+g_{s,t,-}-V_{s,t}-\ell_{s,t}=0,\qquad
    \mbox{on $(a,b)$,}
\]
so that by the identity theorem,
\begin{equation}\label{property phi: 2}
    2g_{s,t}-V_{s,t}-\ell_{s,t}=-2\phi_{s,t},\qquad \mbox{on $\mathcal{V}\setminus(-\infty,a]$.}
\end{equation}
Using (\ref{property g: 3}) this yields,
\begin{align}
    \nonumber
    g_{s,t,+}+g_{s,t,-}-V_{s,t}-\ell_{s,t}
    &=2g_{s,t,-}-V_{s,t}-\ell_{s,t}+
    (g_{s,t,+}-g_{s,t,-})\\[1ex]&=-2\phi_{s,t,-}+2\pi
    i,& \mbox{on $(-\infty,a)$}.\label{property phi: 3}
\end{align}

Inserting (\ref{equation-phig-bulk}), (\ref{property phi: 2}), and
(\ref{property phi: 3}) into (\ref{RHP T: b}), the jump matrix for
$T$ can be written in terms of $\phi_{s,t}$ as
\begin{equation}
    v_T=
    \begin{cases}
        \begin{pmatrix}
            e^{2n\phi_{s,t,+}} & 1 \\
                    0 & e^{2n\phi_{s,t,-}}
                \end{pmatrix}, & \mbox{on $(a,b)$,} \\[3ex]
                \begin{pmatrix}
                    1 & e^{-2n\phi_{s,t,-}} \\
                    0 & 1
                \end{pmatrix}, & \mbox{on $\mathbb R\setminus (a,b)$.}
    \end{cases}
\end{equation}

%
%
%
%
%
%
%

\medskip

It is straightforward to check, using the fact that
$\phi_{s,t,+}+\phi_{s,t,-}=0$ on $(a,b)$, that $v_T$ has on the
interval $(a,b)$ the following factorization,
\begin{equation}\label{factorization}
    v_T
    =
    \begin{pmatrix}
        1 & 0 \\
        e^{2n \phi_{s,t,-}} & 1
    \end{pmatrix}
    \begin{pmatrix}
        0 & 1 \\
        -1 & 0
    \end{pmatrix}
    \begin{pmatrix}
        1 & 0 \\
        e^{2n\phi_{s,t,+}} & 1
    \end{pmatrix},\qquad \mbox{on $(a,b)$,}
\end{equation}
and the opening of the lens is based on this factorization. Observe
that, since $\Re \phi_{s,t,\pm}(x)=0$ and
$\Im\phi_{s,t,\pm}(x)=\mp\int_x^bd\nu_{s,t}(u)$ for $x\in(a,b)$ (see
(\ref{equation-phig-bulk})), and since $\nu_{s,t}$ is positive on
$(a+\delta,b-\delta)$ for $\delta>0$ and $s,t$ sufficiently small
(see Remark \ref{remark: nust positive}), it follows (as in
\cite{Deift}) from the Cauchy-Riemann conditions that
\begin{equation}\label{Re phist<0}
    \Re\phi_{s,t}(z)<0,\qquad\mbox{for $|\Im z|\neq 0$ small and $a+\delta<\Re z
    <b-\delta$.}
\end{equation}

\medskip

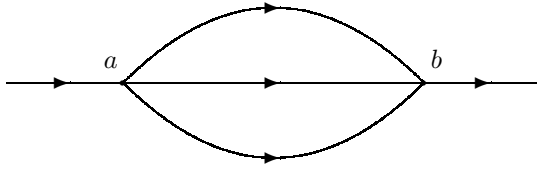
\begin{figure}[t]
\begin{center}
    \setlength{\unitlength}{1mm}
    \begin{picture}(137.5,26)(-2.5,11.5)

        \put(45,25){\thicklines\circle*{.8}} \put(42.5,27){\small $a$}
        \put(85,25){\thicklines\circle*{.8}} \put(86,27){\small $b$}

        \put(94,25){\thicklines\vector(1,0){.0001}}
        \put(29.6,25){\line(1,0){70.4}} \put(38,25){\thicklines\vector(1,0){.0001}}
        \put(66,25){\thicklines\vector(1,0){.0001}}

        \qbezier(45,25)(65,45)(85,25) \put(66,35){\thicklines\vector(1,0){.0001}}
        \qbezier(45,25)(65,5)(85,25) \put(66,15){\thicklines\vector(1,0){.0001}}

    \end{picture}
    \caption{The lens $\Sigma$}
    \label{figure: opening of the lens}
\end{center}
\end{figure}

We deform the RH problem for $T$ into a RH problem for $S$ by
opening a lens as shown in Figure \ref{figure: opening of the lens},
so that we obtain a contour $\Sigma$. For now, we choose the lens to
be contained in $\mathcal V$, but we will specify later how we
choose the lens exactly. Let
\begin{equation}\label{SinT}
    S(z)=
    \begin{cases}
        T(z), & \mbox{for $z$ outside the lens.} \\[1ex]
        T(z)
            \begin{pmatrix}
                1 & 0\\
                -e^{2n\phi_{s,t}(z)} & 1
            \end{pmatrix}, & \mbox{for $z$ in the upper part of the lens,}\\[3ex]
        T(z)
            \begin{pmatrix}
                1 & 0 \\
                e^{2n\phi_{s,t}(z)}& 1
            \end{pmatrix}, & \mbox{for $z$ in the lower part of the lens.}
    \end{cases}
\end{equation}
Then, using (\ref{factorization}) and the RH conditions for $T$,
one can check that $S$ is the unique solution of the following RH
problem.

\subsubsection*{RH problem for $S$:}
\begin{itemize}
    \item[(a)] $S:\mathbb{C}\setminus\Sigma\to\mathbb{C}^{2\times 2}$ is analytic.
    \item[(b)] $S_+(z)=S_-(z)v_S(z)$ for $z\in\Sigma$, with
        \begin{equation}\label{RHP S: b}
            v_S=
            \begin{cases}
                \begin{pmatrix}
                    0 & 1 \\
                    -1 & 0 \\
                \end{pmatrix}, & \mbox{on $(a,b)$,}
                \\[3ex]
                \begin{pmatrix}
                    1 & 0 \\
                    e^{2n \phi_{s,t}} & 1 \\
                \end{pmatrix}, & \mbox{on $\Sigma\cap\mathbb C_\pm$,}
                \\[3ex]
                \begin{pmatrix}
                    1 & e^{-2n\phi_{s,t,-}} \\
                    0 & 1
                \end{pmatrix}, & \mbox{on $\mathbb R\setminus(a,b)$.}
            \end{cases}
        \end{equation}
    \item[(c)] $S(z)=I+\bigO(1/z)$, \qquad as $z\to\infty$.
    \end{itemize}

\begin{remark}\label{remark: exponential decay on lips}
    On the lips of the lens (away from $a$ and $b$) and on
    $\mathbb R\setminus[a-\delta,b+\delta]$, it follows from (\ref{Re phist<0}) and
    (\ref{nust: variational inequality}) that
    the jump matrix for $S$ converges exponentially fast to the identity
    matrix as $n\to\infty$. This convergence is uniform as long as we stay
    away from small disks surrounding the endpoints $a$ and $b$. Near these
    endpoints we have to construct local parametrices.
\end{remark}

\subsection{Parametrix $P^{(\infty)}$ for the outside region}
    \label{subsection: sd-outside}

From Remark \ref{remark: exponential decay on lips}, we expect that
the leading order asymptotics of $Y$ will be determined by a
solution $P^{(\infty)}$ of the following RH problem.

\subsubsection*{RH problem for $P^{(\infty)}$:}
\begin{itemize}
    \item[(a)] $P^{(\infty)}:\mathbb{C}\setminus [a,b]\to\mathbb{C}^{2\times 2}$ is
        analytic.
    \item[(b)] $P^{(\infty)}_+(x)=P^{(\infty)}_-(x)
        \begin{pmatrix}
            0 & 1 \\
            -1 & 0
        \end{pmatrix}$,\qquad for $x\in(a,b)$.
    \item[(c)] $P^{(\infty)}(z)=I+\bigO(1/z)$,\qquad as $z\to\infty$.
\end{itemize}

\medskip

It is well known, see for example \cite{Deift,DKMVZ1}, that
$P^{(\infty)}$ given by
\begin{equation}\label{definition: Pinfty}
    P^{(\infty)}(z)=\begin{pmatrix}1&1\\i&-i\end{pmatrix}
    \left(\frac{z-b}{z-a}\right)^{\sigma_3 /4}
    \begin{pmatrix}1&1\\i&-i\end{pmatrix}^{-1},\qquad\mbox{for $z\in\mathbb C\setminus[a,b]$,}
\end{equation}
is a solution to the above RH problem. Note that $P^{(\infty)}$ is
independent of the parameters $s,t$ and $n$.

\subsection{Parametrix $P^{(a)}$ near the regular endpoint $a$}
    \label{subsection: parametrix near a}

Here, we do the local analysis near the regular endpoint $a$. Let
$U_{\delta,a}=\{z\in\mathbb{C}:|z-a|<\delta\}$ be a small disk with
center $a$ and radius $\delta>0$ sufficiently small such that the
disk lies in $\mathcal V$. We seek a $2\times 2$ matrix valued
function $P^{(a)}$ (depending on the parameters $n,s,$ and $t$) in
the disk $U_{\delta,a}$ with the same jumps as $S$ and which matches
with $P^{(\infty)}$ on the boundary $\partial U_{\delta,a}$ of the
disk. We thus seek a $2\times 2$ matrix valued function that
satisfies the following RH problem.

\subsubsection*{RH problem for $P^{(a)}$:}

\begin{itemize}
    \item[(a)] $P^{(a)}:U_{\delta,a}\setminus\Sigma\to\mathbb C^{2\times 2}$ is analytic.
    \item[(b)] $P^{(a)}_+(z)=P^{(a)}_-(z)v_S(z)$ for $z\in \Sigma\cap U_{\delta,a}$,
        where $v_S$ is given by (\ref{RHP S: b}).
    \item[(c)] $P^{(a)}$ satisfies the matching condition
        \begin{equation}
            P^{(a)}(z)(P^{(\infty)})^{(-1)}(z)=I+\bigO(n^{-1/7}),
        \end{equation}
        as $n\to\infty$ and $s,t\to 0$ such that (\ref{lim stn-1}) holds, uniformly for
        $z\in\partial U_{\delta,a}\setminus\Sigma$.
\end{itemize}

\subsubsection{Airy model RH problem}

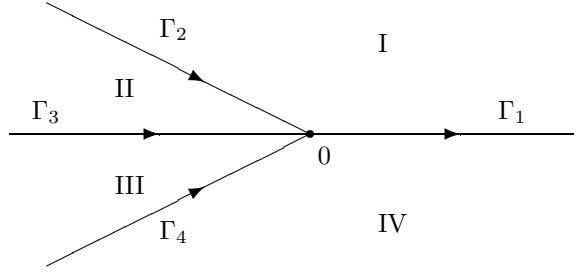
\begin{figure}[t]
    \begin{center}
    \setlength{\unitlength}{1mm}
    \begin{picture}(95,47)(0,2)
        \put(30,38){\small $\Gamma_2$}
        \put(13,27){\small $\Gamma_3$}
        \put(30,11){\small $\Gamma_4$}
        \put(75,27){\small $\Gamma_1$}

        \put(59,36){\small $\I$}
        \put(24,30){\small $\II$}
        \put(24,17){\small $\III$}
        \put(59,12){\small $\IV$}

        \put(50,25){\thicklines\circle*{.9}}
        \put(51,21){\small 0}

        \put(50,25){\line(-2,1){35}} \put(36,32){\thicklines\vector(2,-1){.0001}}
        \put(50,25){\line(-2,-1){35}} \put(36,18){\thicklines\vector(2,1){.0001}}
        \put(50,25){\line(-1,0){40}} \put(30,25){\thicklines\vector(1,0){.0001}}
        \put(50,25){\line(1,0){35}} \put(70,25){\thicklines\vector(1,0){.0001}}
    \end{picture}
    \caption{The oriented contour $\Gamma$. The four
        straight rays $\Gamma_1,\ldots ,\Gamma_4$ divide
        the complex plane into four regions $\I,\II, \III$ and $\IV$.}
        \label{figure: contour gamma}
    \end{center}
\end{figure}

We will construct $P^{(a)}$ by introducing an auxiliary $2\times 2$
matrix valued function $A(\zeta;r)$ with jumps (in the variable
$\zeta$) on an oriented contour $\Gamma=\bigcup_j\Gamma_j$, shown in
Figure \ref{figure: contour gamma}, consisting of four straight rays
\[
    \Gamma_1:\arg\zeta=0,\qquad\Gamma_2:\arg\zeta=\frac{6\pi}{7},
    \qquad\Gamma_3:\arg\zeta=\pi,\qquad\Gamma_4:\arg\zeta=-\frac{6\pi}{7}.
\]
These four rays divide the complex plane into four regions
$\I,\II,\III,$ and $\IV$, also shown in Figure \ref{figure: contour
gamma}. Put
\[
    y_j=y_j(\zeta;r)=\omega^j\Ai(\omega^j(\zeta+r)),\qquad j=0,1,2,
\]
with $\omega=e^{\frac{2\pi i}{3}}$ and with $\Ai$ the Airy function,
and let,
\begin{equation}
    A(\zeta;r)=\sqrt{2\pi}e^{-\frac{\pi i}{4}}\times
    \begin{cases}
        \begin{pmatrix}
            y_0 & -y_2\\
            y_0' & -y_2'
        \end{pmatrix}, & \quad\mbox{for $\zeta\in\I$,}
        \\[3ex]
        \begin{pmatrix}
            -y_1 & -y_2\\
            -y_1' & -y_2'
        \end{pmatrix}, & \quad\mbox{for $\zeta\in\II$,}
        \\[3ex]
        \begin{pmatrix}
            -y_2 & y_1\\
            -y_2' & y_1'
        \end{pmatrix}, & \quad\mbox{for $\zeta\in\III$,}
        \\[3ex]
        \begin{pmatrix}
            y_0 & y_1\\
            y_0' & y_1'
        \end{pmatrix}, & \quad\mbox{for $\zeta\in\IV$.}
    \end{cases}
\end{equation}
With $y_j'$ we mean the derivative of $y_j$ with respect to $\zeta$.
It is well-known, see e.g.\ \cite{Deift,DKMVZ1}, that $A$ satisfies
the following RH problem.

\subsubsection*{RH problem for $A$:}

\begin{itemize}
    \item[(a)] $A$ is analytic for $\zeta\in\mathbb{C}\setminus\Gamma$
        and for $r$ in $\mathbb C$.
    \item[(b)] $A$ satisfies the following jump relations on $\Gamma$,
        \begin{align}
            \label{jumps A: eq1}
            & A_+(\zeta)=A_-(\zeta)
                \begin{pmatrix}
                    0 & 1 \\
                    -1 & 0
                \end{pmatrix}, &\mbox{for $\zeta\in\Gamma_3$,}
            \\[1ex]
            \label{jumps A: eq2}
            & A_+(\zeta)=A_-(\zeta)
                \begin{pmatrix}
                    1 & 1 \\
                    0 & 1
                \end{pmatrix}, &\mbox{for $\zeta\in\Gamma_1$,}
            \\[1ex]
            \label{jumps A: eq3}
            & A_+(\zeta)=A_-(\zeta)
                \begin{pmatrix}
                    1 & 0 \\
                    1 & 1
                \end{pmatrix}, &\mbox{for $\zeta\in\Gamma_2\cup\Gamma_4$.}
        \end{align}
    \item[(c)] $A$ has the following asymptotic behavior at infinity,
        \begin{align}\label{RHP:A-c}
            \nonumber
            A(\zeta;r) &= (\zeta+r)^{-\frac{\sigma_3}{4}}N
                \left[I+\bigO\left((\zeta+r)^{-3/2}\right)\right]
                e^{-\frac{2}{3}(\zeta+r)^{3/2}\sigma_3}
            \\[2ex]
            \nonumber
            &= \zeta^{-\frac{\sigma_3}{4}}N
                \left[I-\frac{1}{4}r^2\zeta^{-1/2}\sigma_3+
                \frac{1}{32}r^4\zeta^{-1}I+\bigO(r^6\zeta^{-3/2})+
                \bigO(r\zeta^{-1})\right]
            \\[1ex]
            &\qquad\qquad\times\, e^{-(\frac{2}{3}\zeta^{3/2}+r\zeta^{1/2})\sigma_3},
        \end{align}
        as $\zeta\to\infty$, uniformly for $r$ such that
        \begin{equation}\label{RHP:A-c2}
            \sgn\bigl(\Im(\zeta+r)\bigr)=\sgn\bigl(\Im \zeta\bigr),
                \qquad\mbox{and}\qquad |r|<|\zeta|^{1/4}.
        \end{equation}
        In (\ref{RHP:A-c}), $N$ is given by
        \begin{equation}\label{definition: N}
            N=\frac{1}{\sqrt 2}
                \begin{pmatrix}
                    1 & 1 \\
                    -1 & 1
                \end{pmatrix} e^{-\frac{1}{4}\pi i\sigma_3}.
        \end{equation}
\end{itemize}

\subsubsection{Construction of $P^{(a)}$}

We seek $P^{(a)}$ in the following form
\begin{equation}\label{definition: Pregular}
    P^{(a)}(z)=E^{(a)}(z)\sigma_3
        A\left(n^{2/3}f_a(z);n^{2/3}r_{s,t}(z)\right)\sigma_3
        e^{n\phi_{s,t}(z)\sigma_3},
\end{equation}
where $E^{(a)}$ is an invertible $2\times 2$ matrix valued function
analytic on $U_{\delta,a}$ and where $f_a$ and $r_{s,t}$ are
(scalar) analytic functions on $U_{\delta,a}$ which are real on
$(a-\delta,a+\delta)$. In addition we take $f_a$ to be a conformal
map from $U_{\delta,a}$ onto a convex neighborhood
$f_a(U_{\delta,a})$ of $0$ such that $f_a(a)=0$ and $f_a'(a)<0$.

If those conditions are all satisfied, and if we open the lens
$\Sigma$ (recall that the lens was not yet fully specified) such
that
\[
    f_a(\Sigma\cap (U_{\delta,a}\cap\mathbb{C}_+))=\Gamma_4\cap
    f_a(U_{\delta,a}),
    \qquad\mbox{and}\qquad
    f_a(\Sigma\cap (U_{\delta,a}\cap\mathbb{C}_-))=\Gamma_2\cap
    f_a(U_{\delta,a}),
\]
then it is straightforward to verify, using (\ref{RHP S: b}) and
(\ref{jumps A: eq1})--(\ref{jumps A: eq3}), that $P^{(a)}$ defined
by (\ref{definition: Pregular}) satisfies conditions (a) and (b) of
the RH problem for $P^{(a)}$.

\medskip

Let
\begin{align}\label{definition: fa}
    \nonumber
    f_a(z)&=\left[\frac{3}{2}\left(-\pi
    i\int_z^a\psi_0(\xi)d\xi\right)(a-z)^{-3/2}\right]^{2/3}(a-z)\\[1ex]
    &=-\left(\frac{1}{2}h_0(a)\sqrt{b-a}\right)^{2/3}(z-a)+\bigO((z-a)^2),\qquad\mbox{as
    $z\to a$,}
\end{align}
where we have used (\ref{definition: psi0}), and let
\begin{equation}\label{definition: rst}
    r_{s,t}(z)=\left(-\pi i\int_z^a(s\psi_1(\xi)+t\psi_2(\xi))d\xi\right) f_a(z)^{-1/2}.
\end{equation}
Then, $f_a$ is analytic with $f_a(a)=0$ and $f_a'(a)<0$, it is real
on $(a-\delta,a+\delta)$, and it is a conformal mapping on
$U_{\delta,a}$ provided $\delta>0$ is sufficiently small. Further,
it is straightforward to check that $r_{s,t}$ is analytic on
$U_{\delta,a}$ and real on $(a-\delta,a+\delta)$, as well. Thus,
$f_a$ and $r_{s,t}$ satisfy the above conditions, so that $P^{(a)}$
defined by (\ref{definition: Pregular}), with $E^{(a)}$ any
invertible analytic matrix valued function, satisfies conditions (a)
and (b) of the RH problem for $P^{(a)}$.

\begin{remark}\label{remark: choice of fa rst}
    We can use any functions $f_a$ and
    $r_{s,t}$, satisfying the conditions stated under equation
    (\ref{definition: Pregular}), to construct the parametrix $P^{(a)}$. However, we
    have to choose them so as to compensate for the factor
    $e^{n\phi_{s,t}\sigma_3}$ in (\ref{definition: Pregular}). Using
    (\ref{definition: psist}), (\ref{definition: phist}), and the fact
    that $\int_a^b\psi_{s,t+}(u)du=1$ we have
    \begin{equation}\label{important relation f and rst}
        e^{-n\left(\frac{2}{3}f_a(z)^{3/2}+r_{s,t}(z)f_a(z)^{1/2}\right)\sigma_3}
            = (-1)^n e^{-n\phi_{s,t}(z)\sigma_3},\qquad \mbox{for $z\in U_{\delta,a}\setminus[a,a+\delta)$.}
    \end{equation}
    From this and (\ref{RHP:A-c}) it is clear that
    our choice of $f_a$ and $r_{s,t}$ will do the job.
\end{remark}

It now remains to determine $E^{(a)}$ such that the matching
condition (c) holds as well. In order to do this we make use of the
following result.

\begin{proposition}\label{proposition: asymptotic expansion Pregular}
    Let $n\to\infty$ and $s,t\to 0$ such that {\rm (\ref{lim
    stn-1})} holds. Then,
    \begin{multline}\label{asymptotic expansion: Pregular}
        P^{(a)}(z)=(-1)^n E^{(a)}(z)
        \left(n^{2/3}f_a(z)\right)^{-\sigma_3/4}\sigma_3 N\sigma_3
        \\[1ex]
        \times\,\left[I-\frac{(n^{4/7}r_{s,t}(z))^2}{4f_a(z)^{1/2}}\, \sigma_3\, n^{-1/7}
        +\frac{(n^{4/7}r_{s,t}(z))^4}{32f_a(z)}\, I\,
        n^{-2/7}+\bigO(n^{-3/7})\right].
    \end{multline}
\end{proposition}

\begin{proof}
    We will use the asymptotics (\ref{RHP:A-c}) of $A$. In order to
    do this we have to check that condition (\ref{RHP:A-c2}) is satisfied
    for our choice of $\zeta=n^{2/3}f_a(z)$ and
    $r=n^{2/3}r_{s,t}(z)$.

    Obviously $r_{s,t}(z)=\bigO(n^{-4/7})$ as $n\to\infty$ and $s,t\to 0$ such
    that (\ref{lim stn-1}) holds, uniformly for $z\in
    \partial U_{\delta,a}$. Then, it is straightforward to check that
    there exists $n_0\in\mathbb{N}$ sufficiently large, and
    $\kappa_1,\kappa_2>0$ sufficiently small, such that
    \begin{equation}\label{asymptotics A: property 1 checked}
        |n^{2/3}r_{s,t}(z)|<|n^{2/3}f_a(z)|^{1/4},
    \end{equation}
    for $z\in\partial U_{\delta,a}$ (for a possible smaller $\delta$),
    for $n\geq n_0$, and for $s$ and $t$ such that $|c_1
    n^{6/7}s-s_0|\leq \kappa_1$ and $|c_1 n^{4/7}t-t_0|\leq \kappa_2$.

    Further, since $f_a$ and $r_{s,t}$ are analytic near $a$ and real
    valued on $(a-\delta,a+\delta)$ one can check that
    \begin{align*}
        &\Im f_a(z)=f_a'(\Re z)\Im z+\bigO((\Im z)^2),&\mbox{as $z\to
        a$,}
        \\[1ex]
        & \Im r_{s,t}(z)=r_{s,t}'(\Re z)+\bigO((\Im z)^2),&\mbox{as $z\to
        a$.}
    \end{align*}
    Now, since $f_a'(a)\neq 0$ and $r_{s,t}'(\Re z)=\bigO(n^{-4/7})$
    uniformly for $z\in U_{\delta,a}$ one then can find a constant $C>0$
    such that,
    \[
        |\Im r_{s,t}(z)|<C|\Im z|<|\Im f_a(z)|,
    \]
    for $z\in\partial U_{\delta,a}\setminus(a-\delta,a+\delta)$, for
    $n\geq n_0$, and for $s$ and $t$ such that $|c_1 n^{6/7}s-s_0|\leq
    \kappa_1$ and $|c_1 n^{4/7}t-t_0|\leq \kappa_2$ (for a possible
    smaller $\delta,\kappa_1$ and $\kappa_2$, and for a possible larger
    $n_0$). This yields
    \begin{equation}\label{asymptotics A: property 2 checked}
        \sgn(\Im (f_a(z)+r_{s,t}(z)))=\sgn(\Im f_a(z)).
    \end{equation}

    We now have shown that condition (\ref{RHP:A-c2}) is satisfied
    so that we can use the asymptotic behavior (\ref{RHP:A-c}) of
    $A$. Using (\ref{definition: Pregular}), (\ref{RHP:A-c}), (\ref{important relation f and
    rst}),
    and the fact that $r_{s,t}(z)=\bigO(n^{-4/7})$ we obtain (\ref{asymptotic expansion:
    Pregular}).
\end{proof}

From (\ref{asymptotic expansion: Pregular}) and the fact that
$r_{s,t}=\bigO(n^{-4/7})$ it is clear that (in order that the
matching condition (c) is satisfied) we have to define $E^{(a)}$ by,
\begin{equation}\label{definition: Ea}
    E^{(a)}=(-1)^n P^{(\infty)}\sigma_3 N^{-1}\sigma_3
    (n^{2/3}f_a)^{\sigma_3/4}.
\end{equation}
Obviously, $E^{(a)}$ is well-defined and analytic in
$U_{\delta,a}\setminus(a,a+\delta)$. Further, using condition (b) of
the RH problem for $P^{(\infty)}$, equation (\ref{definition: N}),
and the fact that $f_{a,-}^{1/4}=if_{a,+}^{1/4}$ on $(a,a+\delta)$,
it is straightforward to check that $E^{(a)}$ has no jump on
$(a,a+\delta)$. We then have that $E^{(a)}$ is analytic in
$U_{\delta,a}$ except for a possible isolated singularity at $a$.
However, $E^{(a)}$ has at most a square root singularity at $a$ and
hence it has to be a removable singularity. Further, since $\det
P^{(\infty)}\equiv 1$ and $\det N=1$ it is clear that $\det
E^{(a)}\equiv 1$ and thus $E^{(a)}$ is invertible. This ends the
construction of the parametrix near the regular endpoint.

\subsection{Parametrix $P^{(b)}$ near the critical endpoint $b$}
    \label{subsection: parametrix near b}

Here, we do the local analysis near the critical endpoint $b$. Let
$U_{\delta,b}=\{z\in\mathbb{C}:|z-b|<\delta\}$ be a small disk with
center $b$ and radius $\delta>0$ sufficiently small such
$U_{\delta,b}$ lies in $\mathcal V$ and such that the disks
$U_{\delta,a}$ and $U_{\delta,b}$ do not intersect. We seek a
$2\times 2$ matrix valued function $P^{(b)}$ (depending on $n,s$ and
$t$) in the disk $U_{\delta,b}$ with the same jumps as $S$ and with
matches with $P^{(\infty)}$ on the boundary $\partial U_{\delta,b}$
of the disk. We thus seek a $2\times 2$ matrix valued function that
satisfies the following RH problem.

\subsubsection*{RH problem for $P^{(b)}$:}

\begin{itemize}
    \item[(a)] $P^{(b)}:U_{\delta,b}\setminus\Sigma\to\mathbb{C}^{2\times 2}$ is
        analytic.
    \item[(b)] $P^{(b)}_+(z)=P^{(b)}_-(z)v_S(z)$ for $z\in U_{\delta,b}\cap
        \Sigma$, where $v_S$ is given by (\ref{RHP S: b}).
    \item[(c)] $P^{(b)}$ satisfies the matching condition
        \begin{equation}
            P^{(b)}(z)(P^{(\infty)})^{-1}(z)=I+\bigO(n^{-1/7}),
        \end{equation}
        as $n\to\infty$ and $s,t\to 0$ such that (\ref{lim stn-1}) holds, uniformly for
        $z\in\partial U_{\delta,b}\setminus\Sigma$.
\end{itemize}
Due to the singular behavior of the equilibrium measure $d\nu_0(x)$
near $b$, see Assumptions \ref{assumptions} (ii), the Airy
parametrix does not fit near $b$. Instead we use a different model
RH problem associated with the $P_I^2$ equation (\ref{PI2}).

\subsubsection{Model RH problem for the $P_I^2$ equation}

We construct $P^{(b)}$ by introducing the following model RH problem
for the special solution $y$ of the $P_I^2$ equation (\ref{PI2}) as
discussed in Section \ref{subsection: PI2 equation}. This RH problem
depends on two complex parameters $s,t$ and has jumps on the
oriented contour $\Gamma$ as defined in Section \ref{subsection:
parametrix near a}, see Figure \ref{figure: contour gamma}. We seek
a $2\times 2$ matrix valued function $\Psi(\zeta)=\Psi(\zeta;s,t)$
satisfying the following conditions.

\subsubsection*{RH problem for $\Psi$:}

\begin{itemize}
    \item[(a)] $\Psi$ is analytic for $\zeta\in\mathbb{C}\setminus\Gamma$.
    \item[(b)] $\Psi$ satisfies the following jump relations on
    $\Gamma$,
    \begin{align}
        \label{RHP Psi: b1}
        &\Psi_+(\zeta)=\Psi_-(\zeta)
        \begin{pmatrix}
            0 & 1 \\
            -1 & 0
        \end{pmatrix},& \mbox{for $\zeta\in\Gamma_3$,} \\[1ex]
        \label{RHP Psi: b2}
        &\Psi_+(\zeta)=\Psi_-(\zeta)
        \begin{pmatrix}
            1 & 1 \\
            0 & 1
        \end{pmatrix},& \mbox{for $\zeta\in\Gamma_1$,} \\[1ex]
        \label{RHP Psi: b3}
        &\Psi_+(\zeta)=\Psi_-(\zeta)
        \begin{pmatrix}
            1 & 0 \\
            1 & 1
        \end{pmatrix},& \mbox{for $\zeta\in\Gamma_2\cup \Gamma_4$.}
    \end{align}
    \item[(c)] $\Psi$ has the following behavior at infinity,
    \begin{equation}\label{RHP Psi: c}
        \Psi(\zeta)=\zeta^{-\frac{1}{4}\sigma_3}N\left(I-h\sigma_3\zeta^{-1/2}
        +\frac{1}{2}\begin{pmatrix}h^2 & iy\\-iy &
        h^2\end{pmatrix}\zeta^{-1}+\bigO(\zeta^{-3/2})\right)
        e^{-\theta(\zeta;s,t)\sigma_3},
    \end{equation}
    where $y=y(s,t)$ is the special solution of the $P_I^2$ equation (\ref{PI2})
    as discussed in Section \ref{subsection: PI2 equation}, where $\frac{\partial h}{\partial s}=-y$, where $N$ is
    given by (\ref{definition: N}), and where $\theta$ is given by (\ref{definition: theta}).
\end{itemize}

\begin{remark}
    Note that the only difference between the model RH problem for Airy
    functions and the one for $P_{\textrm{I}}^2$ lies in the asymptotic
    condition (c). In particular, in $\theta$ we have an extra
    factor $\zeta^{7/2}$.
\end{remark}

If we fix $s_0,t_0\in\mathbb R$, it was proven in \cite[Lemma 2.3
and Proposition 2.5]{CV} that there exists a neighborhood $\mathcal
U$ of $s_0$ and a neighborhood $\mathcal W$ of $t_0$ such that the
RH problem for $\Psi$ is (uniquely) solvable for all
$(s,t)\in\mathcal U\times\mathcal W$. Furthermore, for
$(s,t)\in\mathcal U\times \mathcal W$, $\Psi$ is analytic both in
$s$ and $t$, and condition (c) holds uniformly for $(s,t)$ in
compact subsets of $\mathcal U\times \mathcal W$.

In \cite[Section 2.3]{CV}, the authors have shown that the solution
$\Psi$ of the RH problem for $\Psi$ satisfies the Lax pair
(\ref{differential equations})--(\ref{introduction: W}). From
(\ref{asymptotics Phi}), (\ref{RHP Psi: c}), and (\ref{RHP Psi: b3})
we then obtain
\begin{equation}\label{Phi in Psi}
    \begin{pmatrix}
        \Phi_1(\zeta;s,t) \\
        \Phi_2(\zeta;s,t)
    \end{pmatrix}=
    \begin{cases}
    \begin{pmatrix}
        \Psi_{11}(\zeta;s,t) \\
        \Psi_{21}(\zeta;s,t)
    \end{pmatrix},& \quad\mbox{for $0<\Arg\,\zeta<6\pi/7$,}
    \\[3ex]
    \begin{pmatrix}
        \Psi_{11}(\zeta;s,t) \\
        \Psi_{21}(\zeta;s,t)
    \end{pmatrix}
    \begin{pmatrix}
        1 & 0 \\
        1 & 1
    \end{pmatrix}, & \quad\mbox{for $6\pi/7<\Arg\, \zeta<\pi$.}
    \end{cases}
\end{equation}

\subsubsection{Construction of $P^{(b)}$}

We seek $P^{(b)}$ in the following form
\begin{equation}\label{definition: Psingular}
    P^{(b)}(z)=E^{(b)}(z)\Psi\left(n^{2/7}f_b(z);n^{6/7}sf_1(z),n^{4/7} tf_2(z)\right)
        e^{n\phi_{s,t}(z)\sigma_3},
\end{equation}
where $E^{(b)}$ is an invertible $2\times 2$ matrix valued function
analytic on $U_{\delta,b}$ and where $f_b$, $f_1$, and $f_2$ are
(scalar) analytic functions on $U_{\delta,b}$ which are real on
$(b-\delta,b+\delta)$. We take $f_1$ and $f_2$ to be such that
$f_1(b)=c_1$ and $f_2(b)=c_2$ (where $c_1$ and $c_2$ are given by
(\ref{definition: cc1c2})). Then it is clear from (\ref{lim stn-1})
that for $n$ sufficiently large and $s$ and $t$ sufficiently small,
\[
    n^{6/7}sf_1(z)\in\mathcal U,\qquad \mbox{and}\qquad n^{4/7} tf_2(z)\in\mathcal
    W,\qquad \mbox{for $z\in U_{\delta,b}$,}
\]
where $\mathcal U$ and $\mathcal W$ are the neighborhoods of $s_0$ and $t_0$
where $\Psi$ exists. In addition we take $f_b$ to be a conformal map from
$U_{\delta,b}$ onto a convex neighborhood $f_b(U_{\delta,b})$ of $0$ such that
$f_b(b)=0$ and $f_b'(b)>0$.

If those conditions are all satisfied, and if we open the lens
$\Sigma$ (recall that the lens was not yet fully specified near $b$)
such that
\[
    f_b(\Sigma\cap (U_{\delta,b}\cap\mathbb{C}_+))=\Gamma_2\cap
    f_b(U_{\delta,b}),
    \qquad\mbox{and}\qquad
    f_b(\Sigma\cap (U_{\delta,b}\cap\mathbb{C}_-))=\Gamma_4\cap
    f_b(U_{\delta,a}),
\]
then it is straightforward to verify, using (\ref{RHP S: b}) and
(\ref{RHP Psi: b1})--(\ref{RHP Psi: b3}), that $P^{(b)}$ defined by
(\ref{definition: Psingular}) satisfies conditions (a) and (b) of
the RH problem for $P^{(b)}$.

%

\medskip

Let
\begin{equation}\label{definition: fb}
    f_b(z)=\left[105\left(-\pi i \int_z^b\psi_0(\xi)d\xi\right)(z-b)^{-7/2}\right]^{2/7}(z-b)=c(z-b)+\bigO(z-b)^2,
\end{equation}
as $z\to 0$, where
\[
    c=\left(\frac{15}{2}h_0''(b)\sqrt{b-a}\right)^{2/7}.
\]
To get the expansion of $f_b$ near $b$ we have used
(\ref{definition: psi0}) and the facts that $h_0(b)=h_0'(b)=0$ (see
(\ref{h0(a)=0 etc})). Further since $h_0''(b)>0$ we have that $c>0$.
So, we have defined an analytic function $f_b$ with $f_b(b)=0$ and
$f_b'(b)=c>0$, which is real on $(b-\delta,b+\delta)$, and which is
a conformal mapping on $U_{\delta,b}$ provided $\delta>0$ is
sufficiently small.

Next, let $f_1$ and $f_2$ be defined by
\begin{equation}\label{definition: f1 f2}
    f_1(z)=\left(-\pi i \int_z^b
    \psi_1(\xi)d\xi\right)f_b(z)^{-1/2},
    \quad
    f_2(z)=-3\left(-\pi i\int_z^b\psi_2(\xi)d\xi\right)
    f_b(z)^{-3/2}.
\end{equation}
Since $f_b$ is a conformal mapping in $U_{\delta,b}$ it is clear
from (\ref{definition: psij}) and (\ref{definition: R}) that $f_1$
is analytic in $U_{\delta,b}$. To see that $f_2$ is analytic in
$U_{\delta,b}$ as well, we also need to use the extra condition
$h_2(b)=0$ (see (\ref{h2(b)=0})). Further, $f_1$ and $f_2$ are real
on $(b-\delta,b+\delta)$ and one can check that,
\begin{equation}\label{f1(b)=c1 and f2(b)=c2}
    f_1(b)=\frac{h_1(b)}{c^{1/2}(b-a)^{1/2}}=c_1 ,\qquad
    f_2(b)=-\frac{h_2'(b)}{c^{3/2}(b-a)^{1/2}}=c_2.
\end{equation}
Thus, $f_b, f_1$, and $f_2$ satisfy the above conditions, so that
$P^{(b)}$ defined by (\ref{definition: Psingular}), with $E^{(b)}$
any invertible analytic matrix valued function, satisfies conditions
(a) and (b) of the RH problem for $P^{(b)}$.

\begin{remark}
    As in Remark \ref{remark: choice of fa rst} we note that we could have also used
    different functions $f_b, f_1$, and $f_2$. However, we have to
    choose them so as to compensate for the factor
    $e^{n\phi_{s,t}\sigma_3}$ in (\ref{definition: Psingular}).
    Using (\ref{definition: theta}), (\ref{definition: fb}), (\ref{definition: f1 f2}), (\ref{definition: psist}),
    and (\ref{definition: phist}) we have
    \begin{equation}\label{important relation fbf1f2}
        \theta(n^{2/7}f_b(z);n^{6/7}sf_1(z),n^{4/7}tf_2(z))=n\phi_{s,t}(z),
            \qquad\mbox{for $z\in U_{\delta,b}\setminus(b-\delta,b]$.}
    \end{equation}
    From this and (\ref{RHP Psi: c}) it is clear that our choice of $f_b,f_1$, and
    $f_2$ will do the job.
\end{remark}

It now remains to determine $E^{(b)}$ such that the matching
condition (c) holds as well. In order to do this we make use of the
following proposition (the analogon of Proposition \ref{proposition:
asymptotic expansion Pregular}).

\begin{proposition}
    Let $n\to\infty$ and $s,t\to 0$ such that {\rm (\ref{lim
    stn-1})}
    holds. Then,
    \begin{multline}\label{asymptotic expansion: Psingular}
        P^{(b)}(z)=
        E^{(b)}(z)\left(n^{2/7}f_b(z)\right)^{-\sigma_3/4}N
        \\[1ex]
        \times\,\left[
            I-hf_b(z)^{-1/2}\sigma_3 n^{-1/7}+\frac{1}{2}
            \begin{pmatrix}
                h^2 & iy \\
                -iy & h^2
            \end{pmatrix}f_b(z)^{-1}n^{-2/7}+\bigO(n^{-3/7})
        \right].
    \end{multline}
    where we have used for brevity the notation
    \[
        h=h(n^{6/7}sf_1(z),n^{4/7}tf_2(z)),\qquad\mbox{and}\qquad
        y=y(n^{6/7}sf_1(z),n^{4/7}tf_2(z)).
    \]
\end{proposition}

\begin{proof}
    This follows easily from (\ref{definition: Psingular}), (\ref{RHP Psi:
    c}), and (\ref{important relation fbf1f2})
\end{proof}

From (\ref{asymptotic expansion: Psingular}) it is clear that (in
order that the matching condition (c) is satisfied) we have to
define $E^{(b)}$ by,
\begin{equation}\label{definition: Eb}
    E^{(b)}=P^{(\infty)} N^{-1} \left(n^{2/7}f_b\right)^{\sigma_3/4},
\end{equation}
where $N$ is given by (\ref{definition: N}) and where $P^{(\infty)}$
is the parametrix for the outside region, given by (\ref{definition:
Pinfty}). Similarly as we have proven that $E^{(a)}$ is an
invertible analytic matrix valued function in $U_{\delta,a}$, we can
check that $E^{(b)}$ is invertible and analytic in $U_{\delta,b}$.
This completes the construction of the parametrix near the singular
endpoint.

\subsection{Final transformation: $S\mapsto R$}
    \label{subsection: sd-R}

\begin{figure}[t]
\begin{center}
    \setlength{\unitlength}{1mm}
    \begin{picture}(137.5,26)(-2.5,11.5)

        \put(45,25){\thicklines\circle*{.8}} \put(45,25){\circle{15}}
            \put(46.3,31.9){\thicklines\vector(1,0){.0001}}
        \put(85,25){\thicklines\circle*{.8}} \put(85,25){\circle{15}}
            \put(86.3,31.9){\thicklines\vector(1,0){.0001}}

        \put(92,25){\line(1,0){10.5}} \put(99,25){\thicklines\vector(1,0){.0001}}
        \put(27.5,25){\line(1,0){10.5}} \put(34,25){\thicklines\vector(1,0){.0001}}

        \qbezier(49,30.78)(65,41)(81,30.78) \put(66,35.85){\thicklines\vector(1,0){.0001}}
        \qbezier(49,19.22)(65,9)(81,19.22) \put(66,14.15){\thicklines\vector(1,0){.0001}}

        \put(44.3,27){\small $a$}
        \put(84,27){\small $b$}
    \end{picture}
    \caption{The contour $\Sigma_R$ after the third and final
        transformation.}
    \label{figure: system contours R}
\end{center}
\end{figure}
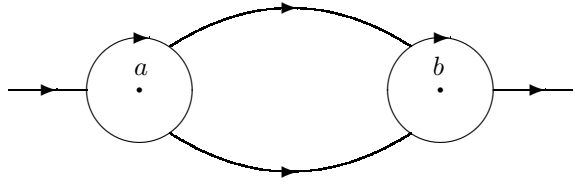

Having the parametrix $P^{(\infty)}$ for the outside region and the
parametrices $P^{(a)}$ and $P^{(b)}$ near the endpoints $a$ and $b$,
we have all the ingredients to perform the final transformation of
the RH problem. Define
\begin{equation}\label{R-S-P}
    R(z)=
    \begin{cases}
        S(z)\left(P^{(a)}\right)^{-1}(z), &\mbox{for $z\in U_{\delta,a}\setminus\Sigma$,} \\[1ex]
        S(z)\left(P^{(b)}\right)^{-1}(z), &\mbox{for $z\in U_{\delta,b}\setminus\Sigma$,} \\[1ex]
        S(z)\left(P^{(\infty)}\right)^{-1}(z), &\mbox{for $z\in
            \mathbb{C}\setminus (\Sigma\cup U_{\delta,a}\cup U_{\delta,b})$.}
    \end{cases}
\end{equation}
Then, by construction of the parametrices, $R$ has only jumps on the
reduced system of contours $\Sigma_R$ shown in Figure \ref{figure:
system contours R}, and $R$ satisfies the following RH problem. The
circles around $a$ and $b$ are oriented clockwise.

\subsubsection*{RH problem for $R$:}

\begin{itemize}
    \item[(a)] $R:\mathbb{C}\setminus\Sigma_R\to \mathbb{C}^{2\times
    2}$ is analytic.
    \item[(b)] $R_+(z)=R_-(z)v_R(z)$ for $z\in\Sigma_R$, with
    \begin{equation}\label{RHP R: b}
        v_R=
        \begin{cases}
            P^{(a)}\left(P^{(\infty)}\right)^{-1}, & \mbox{on $\partial
            U_{\delta,a}$,}\\[1ex]
            P^{(b)}\left(P^{(\infty)}\right)^{-1}, & \mbox{on $\partial
            U_{\delta,b}$,}\\[1ex]
            P^{(\infty)}v_S\left(P^{(\infty)}\right)^{-1}, &
            \mbox{on the rest of $\Sigma_R$.}
        \end{cases}
    \end{equation}
    \item[(c)] $R(z)=I+\bigO(1/z)$,\qquad as $z\to\infty$.
    \item[(d)] $R$ remains bounded near the intersection points of
    $\Sigma_R$.
\end{itemize}

As $n\to\infty$ and $s,t\to 0$ such that (\ref{lim stn-1}) holds, we
have by construction of the parametrices that the jump matrix for
$R$ is close to the identity matrix, both in $L^2$ and $L^{\infty}$-
sense on $\Sigma_R$,
\begin{equation}
    v_R(z)=
    \begin{cases}
        I+\bigO(n^{-1/7}), &\mbox{on $\partial U_{\delta,a}\cup \partial U_{\delta,b}$,} \\[1ex]
        I+\bigO(e^{-\gamma n}),&\mbox{on the rest of $\Sigma_R$,}
    \end{cases}
\end{equation}
with $\gamma>0$ some fixed constant. Then, arguments as in
\cite{DKMVZ2,DKMVZ1} guarantee that $R$ itself is close to the
identity matrix,
\begin{equation}\label{asymptotics R}
    R(z)=I+\bigO(n^{-1/7}),\qquad\mbox{uniformly for $z\in\mathbb{C}\setminus\Sigma_R$,}
\end{equation}
as $n\to\infty$ and $s,t\to 0$ such that (\ref{lim stn-1}) holds.
This completes the Deift/Zhou steepest descent analysis.

\begin{remark}\label{remark: multiple intervals}
    The Deift/Zhou steepest descent method can be generalized to the
    case where the support of $\nu_0$ consists of more than one
    interval. However, there are two (technical) differences.
    First, in the multi-interval case, the equilibrium measures $\nu_1$ and
    $\nu_2$ have densities which are more complicated than in the one-interval
    case, but it remains possible to give explicit formulae.
    Consequently, condition (\ref{conditionV2}), which expresses the
    requirement that the density of $\nu_2$ vanishes at the singular
    endpoint, has to be modified. Further, the construction of the outside parametrix $P^{(\infty)}$
    is more complicated, since it uses
    $\Theta$-functions as in \cite[Lemma 4.3]{DKMVZ2}. With these modifications the
    asymptotic analysis can be carried through in the multi-interval case.

\end{remark}

\subsection{Asymptotics of $R$}
    \label{subsection: sd-asymptoticsR}

For the purpose of proving the universality result for the kernel
$K_n^{(s,t)}$ (Theorem \ref{theorem: universality}) it is enough to
unfold the series of transformations $Y\mapsto T\mapsto S\mapsto R$
and to use (\ref{asymptotics R}). This will be done in the next
section. However, in order to determine the asymptotics of the
recurrence coefficients $a_n^{(n,s,t)}$ and $b_n^{(n,s,t)}$ (Theorem
\ref{theorem: recurrence coefficients}) we need to expand the
$\bigO(n^{-1/7})$ term in (\ref{asymptotics R}).

\bigskip

We show that the jump matrix $v_R$ for $R$ has an expansion of the
form,
\begin{equation}\label{expansion: vR}
    v_R(z)=I+\frac{\Delta_1(z)}{n^{1/7}}+\frac{\Delta_2(z)}{n^{2/7}}+\bigO(n^{-3/7}),
\end{equation}
as $n\to\infty$ and $s,t\to 0$ such that (\ref{lim stn-1}) holds,
uniformly for $z\in\Sigma_R$, and we will explicitly determine
$\Delta_1$ and $\Delta_2$. On $\Sigma_R\setminus(\partial
U_{\delta,a}\cup\partial U_{\delta,b})$, the jump matrix is the
identity matrix plus an exponentially small term, so that
\begin{equation}
    \Delta_1(z)=0,\qquad\Delta_2(z)=0,\qquad\mbox{for $z\in\Sigma_R\setminus(\partial U_{\delta,a}\cup\partial
    U_{\delta,b})$.}
\end{equation}
Now, from (\ref{RHP R: b}), (\ref{asymptotic expansion: Pregular}),
and (\ref{asymptotic expansion: Psingular}) we obtain
(\ref{expansion: vR}) with,
\begin{align}
    \label{definition: Delta1 near a}
    & \Delta_1(z)=-\frac{1}{4}\left(n^{4/7}r_{s,t}(z)\right)^2 f_a(z)^{-1/2}P^{(\infty)}(z)\sigma_3P^{(\infty)}(z)^{-1},
        && \mbox{for $z\in\partial U_{\delta,a}$,}
    \\[2ex]
    \label{definition: Delta1 near b}
    & \Delta_1(z)=
        -h f_0(z)^{-1/2}P^{(\infty)}(z)\sigma_3
        P^{(\infty)}(z)^{-1}, && \mbox{for $z\in\partial U_{\delta,b}$,}
\end{align}
and
\begin{align}
    \label{definition: Delta2 near a}
    & \Delta_2(z)=\frac{1}{32}\left(n^{4/7}r_{s,t}(z)\right)^4
        f_a(z)^{-1} I, &&\mbox{for $z\in\partial U_{\delta,a}$,}
    \\[2ex]
    \label{definition: Delta2 near b}
    & \Delta_2(z)=\frac{1}{2} f_0(z)^{-1}
        \begin{pmatrix}
            h^2 & iy \\
            -iy & h^2
        \end{pmatrix}, &&\mbox{for $z\in\partial U_{\delta,b}$,}
\end{align}
where we have used for brevity
\[
    h=h(n^{6/7}sf_1(z),n^{4/7}tf_2(z)),\qquad
    y=y(n^{6/7}sf_1(z),n^{4/7}tf_2(z)).
\]
    Observe that $\Delta_1$ and $\Delta_2$ have an extension to an
    analytic function in a punctured neighborhood of $a$ and a punctured
    neighborhood of $b$ with simple poles at $a$ and $b$.

    \medskip

    As in \cite[Theorem 7.10]{DKMVZ1} we obtain from (\ref{expansion:
    vR}) that $R$ satisfies,
\begin{equation}\label{expansion: R}
    R(z)=I+\frac{R^{(1)}(z)}{n^{1/7}}+\frac{R^{(2)}(z)}{n^{2/7}}+\bigO(n^{-3/7}),
\end{equation}
as $n\to\infty$ and $s,t\to 0$ such that (\ref{lim stn-1}) holds,
which is valid uniformly for $z\in\mathbb{C}\setminus(\partial
U_{\delta,a}\cup\partial U_{\delta,b})$. We have that
\begin{align}
    \label{condition 1: R(1) R(2)}
    & \mbox{$R^{(1)}$ and $R^{(2)}$ are analytic on $\mathbb C\setminus(\partial
    U_{\delta,a}\cup\partial U_{\delta,b})$,}
    \\[1ex]
    \label{condition 2: R(1) R(2)}
    & R^{(1)}(z)=\bigO(1/z),\qquad R^{(2)}(z)=\bigO(1/z), \qquad \mbox{as $z\to\infty$.}
\end{align}
We will now compute the functions $R^{(1)}$ and $R^{(2)}$
explicitly.

\subsubsection*{Determination of $R^{(1)}$:}

Expanding the jump relation $R_+=R_- v_R$ using (\ref{expansion:
vR}) and (\ref{expansion: R}), and collecting the terms with
$n^{-1/7}$ we find
\[
    R^{(1)}_+(z)=R^{(1)}_-(z)+\Delta_1(z),\qquad\mbox{for $z\in\partial U_{\delta,a}\cup
    \partial U_{\delta,b}$.}
\]
This together with (\ref{condition 1: R(1) R(2)}) and
(\ref{condition 2: R(1) R(2)}) gives an additive RH problem for
$R^{(1)}$. Recall that $\Delta_1$ is analytic in a neighborhood of
$z=a$ and $z=b$ except for simple poles at $a$ and $b$. So,
\[
    \Delta_1(z)=\frac{A^{(1)}}{z-a}+\bigO(1),
        \qquad\mbox{as $z\to a$,}
    \qquad
    \Delta_1(z)=\frac{B^{(1)}}{z-b}+\bigO(1),
        \qquad\mbox{as $z\to b$,}
\]
for certain matrices $A^{(1)}$ and $B^{(1)}$. We then see by
inspection that
\begin{equation}\label{definition: R(1)}
    R^{(1)}(z)=
    \begin{cases}
        {\displaystyle \frac{A^{(1)}}{z-a}+\frac{B^{(1)}}{z-b}}, &\mbox{for $z\in\mathbb C\setminus(\overline{U}_{\delta,a}\cup
        \overline{U}_{\delta,b})$,}
        \\[3ex]
        {\displaystyle \frac{A^{(1)}}{z-a}+\frac{B^{(1)}}{z-b}-\Delta_1(z)},
        &\mbox{for $z\in U_{\delta,a}\cup U_{\delta,b}$,}
    \end{cases}
\end{equation}
solves the additive RH problem for $R^{(1)}$. In now remains to
determine $A^{(1)}$ and $B^{(1)}$. This can be done by expanding the
formulas (\ref{definition: Delta1 near a}) and (\ref{definition:
Delta1 near b}) near $z=a$ and $z=b$, respectively. We then find
after a straightforward calculation (using also the fact that
$f_1(b)=c_1$ and $f_2(b)=c_2$, see (\ref{f1(b)=c1 and f2(b)=c2}),
\begin{align}
        \label{definition: A(1)}
        & A^{(1)}=\frac{1}{8}\sqrt{b-a}\, (n^{4/7}r_{s,t}(a))^2(-f_a'(a))^{-1/2}
            \begin{pmatrix}
                1 & i \\
                i & -1
            \end{pmatrix},
        \\[1ex]
        \label{definition: B(1)}
        & B^{(1)}=\frac{1}{2}h\sqrt{b-a}\, f_b'(b)^{-1/2}
            \begin{pmatrix}
                -1 & i \\
                i & 1
            \end{pmatrix},
\end{align}
where we used $h$ to denote $h(c_1n^{6/7}s,c_2n^{4/7}t)$ for
brevity.

\subsubsection*{Determination of $R^{(2)}$:}

Next, expanding the jump relation $R_+=R_- v_R$ using
(\ref{expansion: vR}) and (\ref{expansion: R}), and collecting the
terms with $n^{-2/7}$ we find
\[
    R^{(2)}_+(z)=R^{(2)}_-(z)+R^{(1)}_-(z)\Delta_1(z)+\Delta_2(z),\qquad
        \mbox{for $z\in\partial U_{\delta,a}\cup
    \partial U_{\delta,b}$.}
\]
This together with (\ref{condition 1: R(1) R(2)}) and
(\ref{condition 2: R(1) R(2)}) gives an additive RH problem for
$R^{(2)}$. Since $R^{(1)}_-$ is the boundary value of the
restriction of $R^{(1)}$ to the disks $U_{\delta,a}$ and
$U_{\delta,b}$ and since $\Delta_1$ and $\Delta_2$ are analytic in a
neighborhood of $a$ and $b$, except for simple poles at $a$ and $b$,
we have
\begin{align*}
    &R^{(1)}(z)\Delta_1(z)+\Delta_2(z)=\frac{A^{(2)}}{z-a}+\bigO(1),\qquad
        \mbox{as $z\to a$,}
    \\[1ex]
    &
    R^{(1)}(z)\Delta_1(z)+\Delta_2(z)=\frac{B^{(2)}}{z-b}+\bigO(1),\qquad
        \mbox{as $z\to b$,}
\end{align*}
for certain matrices $A^{(2)}$ and $B^{(2)}$. As in the
determination of $R^{(1)}$ we then see by inspection that
\begin{equation}\label{definition: R(2)}
    R^{(2)}(z)=
    \begin{cases}
        {\displaystyle \frac{A^{(2)}}{z-a}+\frac{B^{(2)}}{z-b}}, &\mbox{for $z\in\mathbb C\setminus(\overline{U}_{\delta,a}\cup
        \overline{U}_{\delta,b})$,}
        \\[3ex]
        {\displaystyle \frac{A^{(2)}}{z-a}+\frac{B^{(2)}}{z-b}-R^{(1)}(z)\Delta_1(z)-\Delta_2(z)},
        &\mbox{for $z\in U_{\delta,a}\cup U_{\delta,b}$,}
    \end{cases}
\end{equation}
solves the additive RH problem for $R^{(2)}$. The determination of
$A^{(2)}$ and $B^{(2)}$ is more complicated than the determination
of $A^{(1)}$ and $B^{(1)}$. It involves $R^{(1)}(a)$ and
$R^{(1)}(b)$ for which we need to determine also the constant terms
in the expansions of $\Delta_1$ near $z=a$ and $z=b$. After a
straightforward (but rather long calculation) we find,
\begin{align}
    \label{definition: A(2)}
    & A^{(2)}=\frac{(n^{4/7}r_{s,t}(a))^4}{32(-f_a'(a))}
    \begin{pmatrix}
        0 & i \\
        -i & 0
    \end{pmatrix}+\frac{(n^{4/7}r_{s,t}(a))^2 h}{8(-f_a'(a))^{1/2}f_b'(b)^{1/2}}
    \begin{pmatrix}
        1 & i \\
        -i & 1
    \end{pmatrix},
    \\[2ex]
    \label{definition: B(2)}
    & B^{(2)}
    =\frac{y+h^2}{2 f_b'(b)}
    \begin{pmatrix}
        0 & i \\
        -i & 0
    \end{pmatrix}+\frac{(n^{4/7}r_{s,t}(a))^2 h}{8(-f_a'(a))^{1/2}f_b'(b)^{1/2}}
    \begin{pmatrix}
        -1 & i \\
        -i & -1
    \end{pmatrix},
\end{align}
where we used $h$ and $y$ to denote $h(c_1n^{6/7}s,c_2n^{4/7}t)$ and
$y(c_1n^{6/7}s,c_2n^{4/7}t)$ for brevity.

\section{Universality of the double scaling limit}
    \label{section: universality}

Here, we will prove the univarsality result for the 2-point
correlation kernel $K_n^{(s,t)}$. We do this by using the expression
(\ref{KinY}) for $K_n^{(s,t)}$ in terms of $Y$ and by unfolding the
series of transformations $Y\mapsto T\mapsto S\mapsto R$.

\begin{varproof}\textbf{of Theorem \ref{theorem: recurrence coefficients}.}
From equations (\ref{KinY}), (\ref{TinY}), and (\ref{property phi:
2}), the reader can verify that the 2-point kernel $K_n^{(s,t)}$ can
be written as, cf.\ \cite{CK, CKV},
\begin{equation*}
    K_n^{(s,t)}(x,y) = e^{-n\phi_{s,t,+}(x)} e^{-n\phi_{s,t,+}(y)} \frac{1}{2\pi i(x-y)}
        \begin{pmatrix}
            0 & 1
        \end{pmatrix}
        T_+^{-1}(y)T_+(x)
        \begin{pmatrix}
            1 \\
            0
        \end{pmatrix},\quad\mbox{for $x,y\in\mathbb{R}$.}
\end{equation*}
From (\ref{SinT}) and the fact that $S_+=R P^{(b)}_+$ on
$(b-\delta,b+\delta)$, see (\ref{R-S-P}), we have
\[
    T_+=
        \begin{cases}
            R P^{(b)}_+, &\mbox{on $(b,b+\delta)$,}
            \\[1ex]
            R P^{(b)}_+
                \begin{pmatrix}
                    1 & 0 \\
                    e^{2n\phi_{s,t,+}} & 1
                \end{pmatrix}, & \mbox{on $(b-\delta,b)$.}
        \end{cases}
\]
Inserting this in the previous equation for $K_n^{(s,t)}$ we arrive
at,
\begin{equation}\label{KinPhat}
    K_n^{(s,t)}(x,y)
        = e^{-n\phi_{s,t,+}(x)} e^{-n\phi_{s,t,+}(y)} \frac{1}{2\pi i(x-y)}
        \begin{pmatrix}
            0 & 1
        \end{pmatrix}
        \widehat P^{-1}(y)R^{-1}(y)R(x)\widehat P(x)
        \begin{pmatrix}
            1 \\
            0
        \end{pmatrix},
\end{equation}
for $x\in(b-\delta,b+\delta)$, where
\begin{equation}\label{definition: hatP}
    \widehat P=
        \begin{cases}
            P^{(b)}_+,&\mbox{on $(b,b+\delta)$,}
            \\[1ex]
            P^{(b)}_+
            \begin{pmatrix}
                1 & 0 \\
                e^{2n\phi_{s,t,+}} & 1
            \end{pmatrix}, &\mbox{on $(b-\delta,b)$.}
        \end{cases}
\end{equation}
Further, we define
\begin{equation}\label{definition: hatPsi}
    \widehat\Psi=
        \begin{cases}
            \Psi_+&\mbox{on $\mathbb R_+$,}
            \\[1ex]
            \Psi_+
            \begin{pmatrix}
                1 & 0 \\
                1 & 1
            \end{pmatrix},&\mbox{on $\mathbb R_-$,}
        \end{cases}
\end{equation}
where $\Psi$ is the solution of the RH problem for $\Psi$, see
Section \ref{subsection: parametrix near b}. By (\ref{Phi in Psi}),
we have that $\widehat\Psi_{11}=\Phi_1$ and
$\widehat\Psi_{21}=\Phi_2$. Using (\ref{definition: Psingular}),
(\ref{definition: hatP}), and (\ref{definition: hatPsi}) a
straightforward calculation yields,
\[
    \widehat P(x)=
        E^{(b)}(x) \widehat\Psi\left(n^{2/7}f_b(x);n^{6/7}sf_1(x),n^{4/7}tf_2(x)\right)
        e^{n\phi_{s,t,+}(x)\sigma_3},
    \qquad\mbox{for $x\in(b-\delta,b+\delta)$.}
\]
Inserting this into (\ref{KinPhat}) we then obtain,
\begin{multline}\label{KnN: PsiER}
    K_n^{(s,t)}(x,y)=\frac{1}{2\pi i(x-y)}
        \begin{pmatrix}
            0 & 1
        \end{pmatrix}
         \widehat\Psi^{-1}\left(n^{2/7}f_b(y);n^{6/7}sf_1(y),n^{4/7}tf_2(y)\right)
         \\[1ex]
         \times (E^{(b)})^{-1}(y)R^{-1}(y) R(x)E^{(b)}(x)\widehat
        \Psi\left(n^{2/7}f_b(x);n^{6/7}sf_1(x),n^{4/7}tf_2(y)\right)
        \begin{pmatrix}
            1 \\
            0
        \end{pmatrix},
\end{multline}
for $x\in(b-\delta,b+\delta)$.

Now, we introduce for the sake of brevity some notation. Let
\begin{equation}\label{definition: uv}
    u_n= b+\frac{u}{c n^{2/7}},\quad\mbox{and}\quad v_n=b+\frac{v}{cn^{2/7}}, \quad \mbox{with
    $c=f_b'(b)=\left(\frac{15}{2}
    h_0''(b)\sqrt{b-a}\right)^{2/7}$.}
\end{equation}
We then have,
\begin{equation}\label{uv1}
    \lim_{n\to\infty}n^{2/7}f_b(u_n)=u, \qquad\mbox{and}\qquad
    \lim_{n\to\infty}n^{2/7}f_b(v_n)=v.
\end{equation}
Furthermore, since $f_1(b)=c_1$ and $f_2(b)=c_2$ (see (\ref{f1(b)=c1
and f2(b)=c2})) we have in the limit as $n\to\infty$ and $s,t\to 0$
such that (\ref{lim stn-1}) holds,
\begin{align}
    &\lim n^{6/7}sf_1(u_n)=s_0, && \lim n^{6/7}sf_1(v_n)=s_0,\label{uv2}\\
    &\lim n^{4/7}tf_2(u_n)=t_0, && \lim n^{4/7}tf_2(v_n)=t_0.\label{uv3}
\end{align}
Now, a similar argument as in \cite{KV2} shows that
\begin{equation}\label{uv4}
    \lim
    E_b^{-1}(v_n)R(v_n)^{-1}R(u_n)E_b(u_n)=I.
\end{equation}
Inserting (\ref{uv1})--(\ref{uv4}) into (\ref{KnN: PsiER}) and using
the fact that $\widehat\Psi_{11}=\Phi_1$ and
$\widehat\Psi_{21}=\Phi_2$ it is then straightforward to obtain
\begin{align}
    \nonumber
    & \lim \frac{1}{cn^{2/7}}K_n^{(s,t)}(u_n,v_n) \\
    \nonumber
    &\qquad = \frac{1}{2\pi i(u-v)}
    \begin{pmatrix}
            0 & 1
        \end{pmatrix}
        \widehat\Psi^{-1}(v;s_0,t_0)\widehat
        \Psi(u;s_0,t_0)
        \begin{pmatrix}
            1 \\
            0
        \end{pmatrix}\\[1ex]
    &\qquad=\frac{1}{2\pi
i(u-v)}\left(\Phi_1(u;s_0,t_0)\Phi_2(v;s_0,t_0)-\Phi_1(v;s_0,t_0)\Phi_2(u;s_0,t_0)\right),
\end{align}
where we take the limit $n\to\infty$ and $s,t\to 0$ such that
(\ref{lim stn-1}) holds. This completes the proof of Theorem
\ref{theorem: universality}.
\end{varproof}

\section{Asymptotics of the recurrence coefficients}
    \label{section: recurrence}

We will now determine the asymptotics of $a_n^{(n,s,t)}$ and
$b_n^{(n,s,t)}$ as $n\to\infty$ and $s,t\to 0$ such that (\ref{lim
stn-1}) holds. In order to do this, we make use of the following
result, see e.g.\ \cite{Deift,DKMVZ1}. Let $Y$ be the unique
solution of the RH problem for $Y$. There exist $2\times 2$ constant
(independent of $z$ but depending on $n,s$ and $t$) matrices $Y_1$
and $Y_2$ such that
\begin{equation}
    Y(z)\begin{pmatrix}
        z^{-n} & 0 \\
        0 & z^n
    \end{pmatrix}=I+\frac{Y_1}{z}+\frac{Y_2}{z^2}+\bigO(1/z^3),\qquad\mbox{as
    $z\to\infty$,}
\end{equation}
and
\begin{equation}\label{ainY binY}
    a_n^{(n,s,t)}=\sqrt{(Y_1)_{12}(Y_1)_{21}},\qquad
    b_n^{(n,s,t)}=(Y_1)_{11}+\frac{(Y_2)_{12}}{(Y_1)_{12}}.
\end{equation}
We need to determine the constant matrices $Y_1$ and $Y_2$. For
large $|z|$ it follows from (\ref{TinY}), (\ref{SinT}), and
(\ref{R-S-P}), that
\begin{equation}\label{Y outside}
    Y(z)=e^{\frac{1}{2}n\ell_{s,t}\sigma_3}R(z)P^{(\infty)}(z)e^{ng_{s,t}(z)\sigma_3}e^{-\frac{1}{2}n\ell_{s,t}\sigma_3}.
\end{equation}
So, in order to determine $Y_1$ and $Y_2$ we need the asymptotic
behavior of $P^{(\infty)}(z), e^{ng_{s,t}(z)\sigma_3}$, and $R(z)$
as $z\to\infty$.

\subsubsection*{Asymptotic behavior of $P^{(\infty)}(z)$ as $z\to\infty$:}

Expanding the factor $((z-b)/(z-a))^{\sigma_3/4}$ in
(\ref{definition: Pinfty}) at $z=\infty$ it is clear that,
\begin{equation}\label{asymptotics Pinfty}
    P^{(\infty)}(z)=I+\frac{P_1^{(\infty)}}{z}+\frac{P_2^{(\infty)}}{z}+\bigO(1/z^3),\qquad\mbox{as
    $z\to\infty$,}
\end{equation}
with
\begin{equation}\label{P1inftyP2infty}
    P_1^{(\infty)}=\frac{i}{4}(b-a)
        \begin{pmatrix}
            0 & 1 \\
            -1 & 0
        \end{pmatrix},\qquad
    P_2^{(\infty)}=\frac{i}{8}(b^2-a^2)
        \begin{pmatrix}
            * & 1 \\
            -1 & *
        \end{pmatrix}.
\end{equation}

\subsubsection*{Asymptotic behavior of $e^{ng_{s,t}(z)\sigma_3}$ as $z\to\infty$:}

By (\ref{definition: gst}) we have
\begin{equation}\label{asymptotics engst}
    e^{ng_{s,t}(z)\sigma_3}
    \begin{pmatrix}
        z^{-n} & 0 \\
        0 & z^n
    \end{pmatrix}=I+\frac{G_1}{z}+\frac{G_2}{z^2}+\bigO(1/z^3),\qquad\mbox{as $z\to\infty$,}
\end{equation}
with
\begin{equation}\label{definition: G1G2}
    G_1=-n\int_a^b ud\nu_{s,t}(u)
        \begin{pmatrix}
            1 & 0 \\
            0 & -1
        \end{pmatrix},\qquad
    G_2=\begin{pmatrix}
        * & 0 \\
        0 & *
    \end{pmatrix}.
\end{equation}

\subsubsection*{Asymptotic behavior of $R(z)$ as $z\to\infty$:}

As in \cite{DKMVZ1} the matrix valued function $R$ has the following
asymptotic behavior at infinity,
\begin{equation}\label{asymptotics R: R1R2}
    R(z)=I+\frac{R_1}{z}+\frac{R_2}{z^2}+\bigO(1/z^3),\qquad\mbox{as
    $z\to\infty$.}
\end{equation}
The compatibility with (\ref{expansion: R}), (\ref{definition:
R(1)}), and (\ref{definition: R(2)}) yields that
\begin{align}
    \label{definition: R1}
    & R_1=\left(A^{(1)}+B^{(1)}\right)n^{-1/7}+\left(A^{(2)}+B^{(2)}\right)n^{-2/7}+\bigO(n^{-3/7}),
    \\[1ex]
    \label{definition: R2}
    &
    R_2=\left(aA^{(1)}+bB^{(1)}\right)n^{-1/7}+\left(aA^{(2)}+bB^{(2)}\right)n^{-2/7}+\bigO(n^{-3/7}),
\end{align}
as $n\to\infty$ and $s,t\to 0$ such that (\ref{lim stn-1}) holds.
Here, $A^{(1)}$, $B^{(1)}$, $A^{(2)}$, and $B^{(2)}$ are given by
(\ref{definition: A(1)}), (\ref{definition: B(1)}),
(\ref{definition: A(2)}), and (\ref{definition: B(2)}),
respectively.

\medskip

Now, we are ready to determine the asymptotics of the recurrence
coefficients.

\begin{varproof}\textbf{of Theorem \ref{theorem: recurrence coefficients}.}
    Note that by (\ref{Y outside}), (\ref{asymptotics Pinfty}), (\ref{asymptotics engst}) and (\ref{asymptotics R: R1R2}),
    \begin{equation}\label{definition: Y1}
        Y_1= e^{\frac{1}{2}n\ell_{s,t}\sigma_3}\left[P_1^{(\infty)}+G_1+R_1\right]e^{-\frac{1}{2}n\ell_{s,t}\sigma_3}
    \end{equation}
    and
    \begin{equation}\label{definition: Y2}
        Y_2= e^{\frac{1}{2}n\ell_{s,t}\sigma_3}\left[P_2^{(\infty)}+G_2+R_2
            +R_1 P_1^{(\infty)}+\left(P_1^{(\infty)}+R_1\right)G_1\right]e^{-\frac{1}{2}n\ell_{s,t}\sigma_3}
    \end{equation}

    We start with the recurrence coefficient $a_n^{(n,s,t)}$. Inserting
    (\ref{definition: Y1}) into (\ref{ainY binY}), and using the facts that
    $(P_1^{(\infty)})_{12}=-(P_1^{(\infty)})_{21}=i(b-a)/4$ (by (\ref{P1inftyP2infty})) and
    $(G_1)_{12}=(G_1)_{21}=0$ (by (\ref{definition: G1G2})), we
    obtain
    \begin{equation}\label{proof: recurrence: eq1}
        a_n^{(n,s,t)} =
            \left[\left(\frac{b-a}{4}\right)^2+i\frac{b-a}{4}\left((R_1)_{21}
                -(R_1)_{12}\right)+(R_1)_{12}(R_1)_{21}\right]^{1/2}.
    \end{equation}
    Now, from the formula (\ref{definition: R1}) for $R_1$ and the formulas (\ref{definition: A(1)}), (\ref{definition:
    B(1)}), (\ref{definition: A(2)}), and (\ref{definition: B(2)})
    for $A^{(1)}, B^{(1)}, A^{(2)},$ and $B^{(2)}$,
    we have
    \[
        (R_1)_{21} - (R_1)_{12} =
        -i\left[\frac{y}{f_b'(b)}+\left(\frac{(n^{4/7}r_{s,t}(a))^2}{4(-f_a'(a))^{1/2}}
        +\frac{h}{f_b'(b)}\right)^2\right]n^{-2/7}+\bigO(n^{-3/7}),
    \]
    and
    \[
        (R_1)_{12}(R_1)_{21}=-\frac{b-a}{4}\left(\frac{(n^{4/7}r_{s,t}(a))^2}{4(-f_a'(a))^{1/2}}
        +\frac{h}{f_b'(b)}\right)^2
        n^{-2/7}+\bigO(n^{-3/7}).
    \]
    Note that we have used $y$ to denote $y(c_1n^{6/7}s,c_2n^{4/7}t)$ for brevity.
    Inserting the latter two equations into (\ref{proof: recurrence: eq1}) and using
    the fact that $f_b'(b)=c$ (by (\ref{definition: fb})) we then obtain (\ref{theorem: recurrence coefficients: anst}).

    \medskip

    We will now consider the recurrence coefficient $b_n^{(n,s,t)}$.
    Inserting (\ref{definition: Y1}) and (\ref{definition: Y2}) into (\ref{ainY binY}), and
    using the facts that
    $(P_1^{(\infty)})_{11}=(P_1^{(\infty)})_{22}=0$,
    $(G_1)_{12}=(G_2)_{12}=0$, $(G_1)_{11}+(G_1)_{22}=0$, and $R_1=\bigO(n^{-1/7})$, we obtain
    \begin{align*}
        b_n^{(n,s,t)} & =(R_1)_{11}+\frac{(P_2^{(\infty)})_{12}+(R_1)_{11}(P_1^{(\infty)})_{12}+
            (R_2)_{12}}{(P_1^{(\infty)})_{12}+(R_1)_{12}}
        \\[1ex]
        &=(R_1)_{11}+ \left[\frac{(P_2^{(\infty)})_{12}}{(P_1^{(\infty)})_{12}}
        +(R_1)_{11}+\frac{(R_2)_{12}}{(P_1^{(\infty)})_{12}}\right]
        \\[1ex]
        &\qquad\qquad\qquad\qquad\qquad\times\,\left[1-\frac{(R_1)_{12}}{(P_1^{(\infty)})_{12}}+
        \left(\frac{(R_1)_{12}}{(P_1^{(\infty)})_{12}}\right)^2+\bigO(n^{-3/7})\right].
    \end{align*}
    Since $(P_1^{(\infty)})_{12}=i(b-a)/4$,
    $(P_2^{(\infty)})_{12}=i(b^2-a^2)/8$, $R_1=\bigO(n^{-1/7})$, and
    $R_2=\bigO(n^{-1/7})$ we then obtain after a straightforward calculation and combining terms,
    \begin{multline}\label{proof: recurrence: eq2}
        b_n^{(n,s,t)} =\frac{b+a}{2}+\left(2(R_1)_{11}+2i\frac{b+a}{b-a}(R_1)_{12}-\frac{4i}{b-a}(R_2)_{12}\right)
        \left(1+\frac{4i}{b-a}(R_1)_{12}\right)
        \\[1ex]
        -\, \frac{4i}{b-a}(R_1)_{11}
        (R_1)_{12}+\bigO(n^{-3/7}).
    \end{multline}
    Now, from (\ref{definition: R1}), (\ref{definition: R2}), (\ref{definition: A(1)}), (\ref{definition:
    B(1)}), (\ref{definition: A(2)}), and (\ref{definition: B(2)}) we have
    \begin{align*}
        & 2(R_1)_{11}+2i\frac{b+a}{b-a}(R_1)_{12}-\frac{4i}{b-a}(R_2)_{12} =
            2i\left[A^{(2)}_{12}-B^{(2)}_{12}\right]n^{-2/7}+\bigO(n^{-3/7})
        \\[2ex]
        &\qquad\qquad=\left(\frac{y}{f_b'(b)}+\frac{h^2}{f_b'(b)}-\frac{(n^{4/7}r_{s,t}(a))^4}{16(-f_a'(a))}\right)
        n^{-2/7}+\bigO(n^{-3/7}),
    \end{align*}
    and
    \begin{align*}
        (R_1)_{11}(R_1)_{12} &=
        -i\left[\left(A_{12}^{(1)}\right)^2-\left(B_{12}^{(1)}\right)^2\right]n^{-2/7}+\bigO(n^{-3/7})
        \\[2ex]
        &=-i\frac{b-a}{4}\left(\frac{h^2}{f_b'(b)}-\frac{(n^{4/7}r_{s,t}(a))^4}{16(-f_a'(a))}\right)
        n^{-2/7}+\bigO(n^{-3/7}).
    \end{align*}
    Inserting the latter two equations into (\ref{proof: recurrence: eq2}) and using the facts that
    $(R_1)_{12}=\bigO(n^{-1/7})$ and $f_b'(b)=c$ we obtain (\ref{theorem: recurrence coefficients: bnst}).
    So, the theorem is proven.
\end{varproof}

\section*{Acknowledgements}

We thank Arno Kuijlaars for careful reading and stimulating discussions.

The authors are supported by FWO research project G.0455.04, by
K.U.Leuven research grant OT/04/24, and by INTAS Research Network
NeCCA 03-51-6637. The second author is Postdoctoral Fellow of the
Fund for Scientific Research - Flanders (Belgium).

\end{document}